\documentclass[twocolumn,amsmath,amssymb,nofootinbib]{revtex4}
\usepackage{graphicx}
\usepackage{microtype}
\usepackage{epstopdf}
\usepackage{amsmath}
\usepackage{amssymb}
\usepackage{subfigure}
\usepackage{hyperref}
\usepackage{url}
\usepackage{xcolor}
\usepackage{color}
\usepackage{mathrsfs}  
\usepackage{calrsfs}

\definecolor{amaranth}{rgb}{0.9, 0.17, 0.31}
\definecolor{purple(munsell)}{rgb}{0.62, 0.0, 0.77}
\definecolor{americanrose}{rgb}{1.0, 0.01, 0.24}
\definecolor{palatinateblue}{rgb}{0.15, 0.23, 0.89}
\definecolor{royalblue(web)}{rgb}{0.25, 0.41, 0.88}
\definecolor{hanpurple}{rgb}{0.32, 0.09, 0.98}
\definecolor{beaublue}{rgb}{0.74, 0.83, 0.9}
\definecolor{carminered}{rgb}{1.0, 0.0, 0.22}
\definecolor{emerald}{rgb}{0.31, 0.78, 0.47}
\definecolor{vividviolet}{rgb}{0.62, 0.0, 1.0}
\definecolor{brightpink}{rgb}{1.0, 0.0, 0.5}
\hypersetup{ linktoc=all,
    colorlinks, linkcolor={palatinateblue},
    citecolor={brightpink}, urlcolor={amaranth}
}
\newcommand{\changeurlcolor}[1]{\hypersetup{urlcolor=#1}}    
\renewcommand{\d}[1]{\ensuremath{\operatorname{d}\!{#1}}}

\begin{document}

\title{Instanton Tunneling for De Sitter Space with Real Projective Spatial Sections}

\author{Yen Chin Ong}
     \email{ongyenchin@sjtu.edu.cn}
   
     \affiliation{
(1.) Center for Astronomy and Astrophysics, Department of Physics and Astronomy,
Shanghai Jiao Tong University, Shanghai 200240, China\footnote{Present address}. \\
(2.) Nordita, KTH Royal Institute of Technology and Stockholm University, \\ Roslagstullsbacken 23,
SE-106 91 Stockholm, Sweden\\
(3.) Riemann Center for Geometry and Physics, Leibniz Universit\"at Hannover,\\
Appelstrasse 2, 30167 Hannover, Germany\\
}%

\author{Dong-han Yeom}
     \email{innocent.yeom@gmail.com}
   
     \affiliation{Leung Center for Cosmology and Particle Astrophysics, National Taiwan University, Taipei 10617, Taiwan\\
}%

\begin{abstract}
The physics of tunneling from one spacetime to another is often understood in terms of instantons. For some instantons, it was recently shown in the literature that there are two complementary ``interpretations'' for their analytic continuations. Dubbed ``something-to-something'' and ``nothing-to-something'' interpretations, respectively, the former involves situation in which the initial and final hypersurfaces are connected by a Euclidean manifold, whereas the initial and final hypersurfaces in the latter case are not connected in such a way. We consider a de Sitter space with real projective space $\mathbb{R}\text{P}^3$ spatial sections, as was originally understood by de Sitter himself. This original version of de Sitter space has several advantages over the usual de Sitter space with $\text{S}^3$ spatial sections. In particular, the interpretation of the de Sitter entropy as entanglement entropy is much more natural. We discuss the subtleties involved in the tunneling of such a de Sitter space. 
\end{abstract}

\maketitle
\section{de Sitter Space According to\newline de Sitter: The Case for $\mathbb{R}\text{P}^3$}

De Sitter space is a Lorentzian manifold of constant positive curvature, which is a solution to the Einstein's field equations with a positive cosmological constant $\Lambda$. It plays very prominent roles in theoretical cosmology, since the observed accelerating expansion of the universe is very well explained by just such a $\Lambda$ \cite{1002.3966}. Furthermore, de Sitter space also provides the arena for early inflationary universe. Although in cosmology one often uses the flat slicing of de Sitter space, such a slicing does not cover the entire spacetime. Here we are more concerned with the actual, \emph{global}, structure of de Sitter space. Globally, the topology of de Sitter space --- \emph{as we have come to know it} --- is $\mathbb{R} \times \text{S}^3$. However, a little known fact is that de Sitter himself did \emph{not} have the same topology in mind for ``de Sitter space'' \cite{deSitter}. Instead, he preferred to topologically identify the antipodal points on the $\text{S}^3$. This yields the projective space $\mathbb{R}\text{P}^3 \cong \text{S}^3/\mathbb{Z}_2$, which, like $\text{S}^3$, is an orientable spin manifold, and thus is a well-behaved spatial geometry for a spacetime. This is not true for real projective spaces in all dimensions \cite{0308022}.

$\mathbb{R}\text{P}^3$ admits the same metric as that of $\text{S}^3$, since the metric tensor is a local quantity which is not sensitive to the global topology. This is the reason why Einstein's field equations --- a system of differential equations --- cannot uniquely fix the topology. Specifically, a 3-sphere with sectional curvature $1/\ell^2$ admits a metric tensor of the form
\begin{equation}\label{S3}
g(\text{S}^3)=\ell^2 \left[\d\chi^2 + \sin^2\chi \left(\d\theta^2 + \sin^2\theta \d\phi^2\right)\right],
\end{equation}
where $0 \leqslant \chi, \theta \leqslant \pi$, and $0\leqslant \phi < 2\pi$.
For a real projective space, we have \emph{exactly the same} metric expression:
\begin{equation}
g(\mathbb{R}\text{P}^3)=\ell^2 \left[\d\chi^2 + \sin^2\chi \left(\d\theta^2 + \sin^2\theta \d\phi^2\right)\right],
\end{equation}
but now the coordinates on the 3-spheres are mapped in such a way that antipodal points are identified. Namely, $\chi \mapsto \pi - \chi$, $\theta \mapsto \pi-\theta$, and $\phi \mapsto \pi +\phi$. We will refer to this as the \emph{antipodal map}. 

The topology of de Sitter space, as de Sitter understood it, is therefore $\mathbb{R} \times \mathbb{R}\text{P}^3$, which, following \cite{0308022}, we shall denote by $\text{dS}[\mathbb{R}\text{P}^3]$. Similarly, the de Sitter space with $\mathbb{R} \times \text{S}^3$ topology will be denoted by $\text{dS}[\text{S}^3]$. 
The reason de Sitter favoured $\text{dS}[\mathbb{R}\text{P}^3]$ over $\text{dS}[\text{S}^3]$ is that, following Schwarzschild's paper\footnote{Note that this paper of Schwarzschild predated even Special Relativity. For an English translation of the paper, see \cite{Sch}.} in 1900 \cite{Schwarzschild}, the fact that $\text{S}^3$ has antipodal points means that the global geometry ``forces a correlation'' between these two otherwise faraway points. Geometrically, any two coplanar geodesics in $\text{S}^3$ would intersect twice, whereas in $\mathbb{R}\text{P}^3$ they would intersect only once. Since the intersection of geodesics is a matter of \emph{local} physics, it would seem strange that this would somehow force a correlation at the antipodal point, which is well outside the current observable universe! It is for this reason that de Sitter considered $\mathbb{R}\text{P}^3$ to be ``the simpler case, and it is preferable to adopt this for the physical world'' \cite{deSitter}. 

\begin{figure}
\includegraphics[width=3.46in]{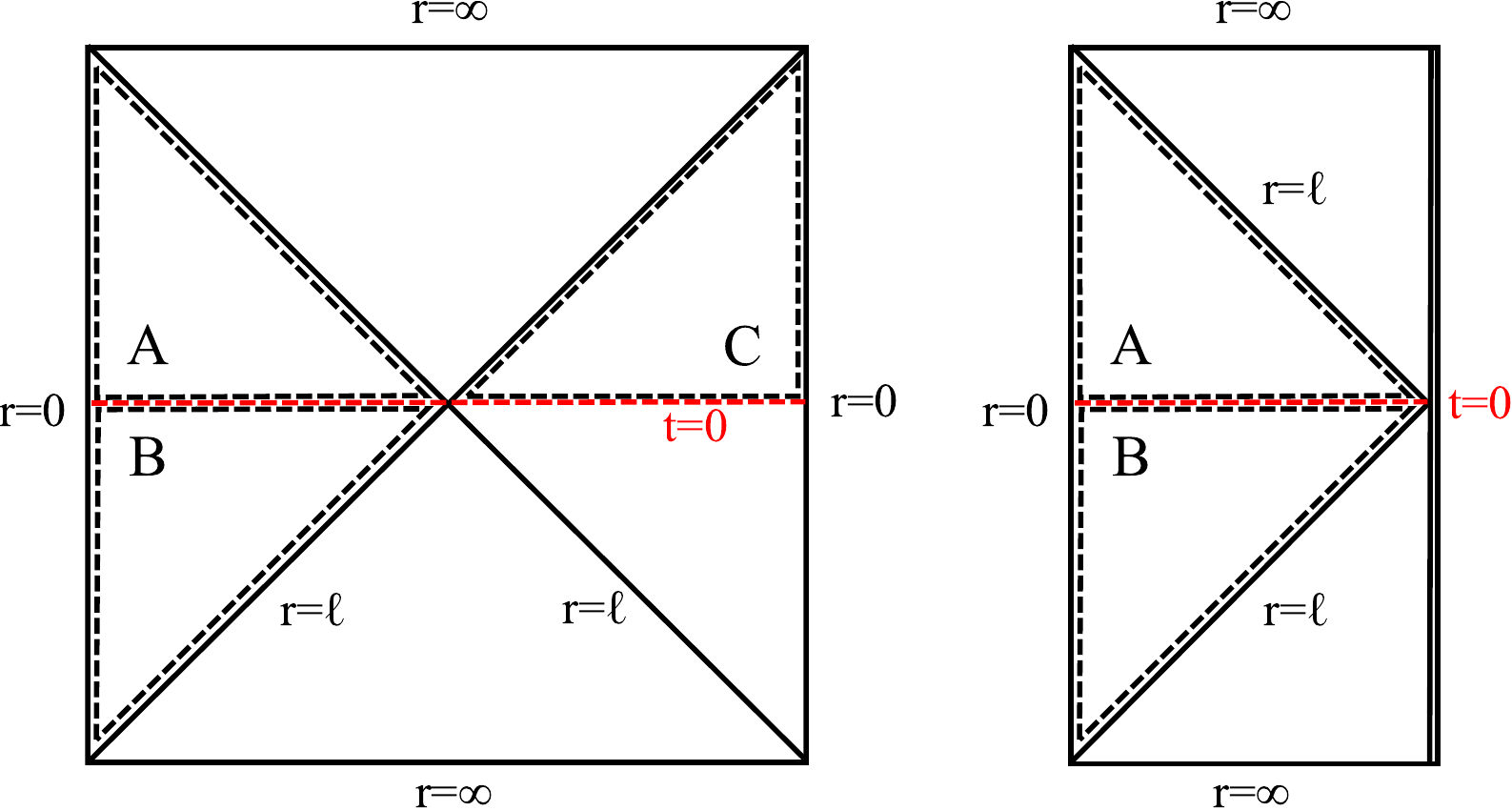} 
\caption{\textbf{Left:} The Penrose diagram of $\text{dS}[\text{S}^3]$; \textbf{Right:} the Penrose diagram of $\text{dS}[\mathbb{R}\text{P}^3]$. Each point of the double line corresponds to a $\mathbb{R}\text{P}^2$ on the equator of $\mathbb{R}\text{P}^3$. Slanted lines are cosmological horizons. The dashed lines should be ignored at this point; they will be discussed later.\label{penrose1}} 
\end{figure}

The Penrose diagram of both $\text{dS}[\text{S}^3]$  and $\text{dS}[\mathbb{R}\text{P}^3]$ are shown in Fig.(\ref{penrose1}). In the familiar case of $\text{dS}[\text{S}^3]$, there are two causal patches that are out of causal contact. An observer centered at the -- arbitrary -- pole $r=0$ would not be able to see the antipode of the $\text{S}^3$ (namely, \emph{the other} $r=0$) at the other end of the universe. In $\text{dS}[\mathbb{R}\text{P}^3]$, due to the topological identification, \emph{there is no antipode}. Since the left and right hemispheres of the $\text{S}^3$ are identified by the antipodal map, only half of the Penrose diagram of $\text{dS}[\text{S}^3]$ remains. The equator of $\text{S}^3$ is a $\text{S}^2$, which, under the antipodal map, becomes a $\mathbb{R}\text{P}^2$. Thus, in the Penrose diagram of $\text{dS}[\mathbb{R}\text{P}^3]$, which is on the right of Fig.(\ref{penrose1}), each point on the double line corresponds to a $\mathbb{R}\text{P}^2$.
An observer, say, Alice, who started at $r=0$ and travels toward the $\mathbb{R}\text{P}^2$ equator, would, upon traveling through $\mathbb{R}\text{P}^2$, find herself now traveling back toward $r=0$, since she is ``reflected'' back by the topological identification. It is, however, not exactly a reflection (since Alice is traveling \emph{through} -- not \emph{within} -- $\mathbb{R}\text{P}^2$). Instead, she would find that the universe seems to have been rotated upside down. See \cite{0308022} for more discussions. Note that this ``reflection'' is not imposed as a boundary condition by hand, instead it is an unavoidable consequence of the topological identification.  

\begin{figure}
\includegraphics[width=3.0in]{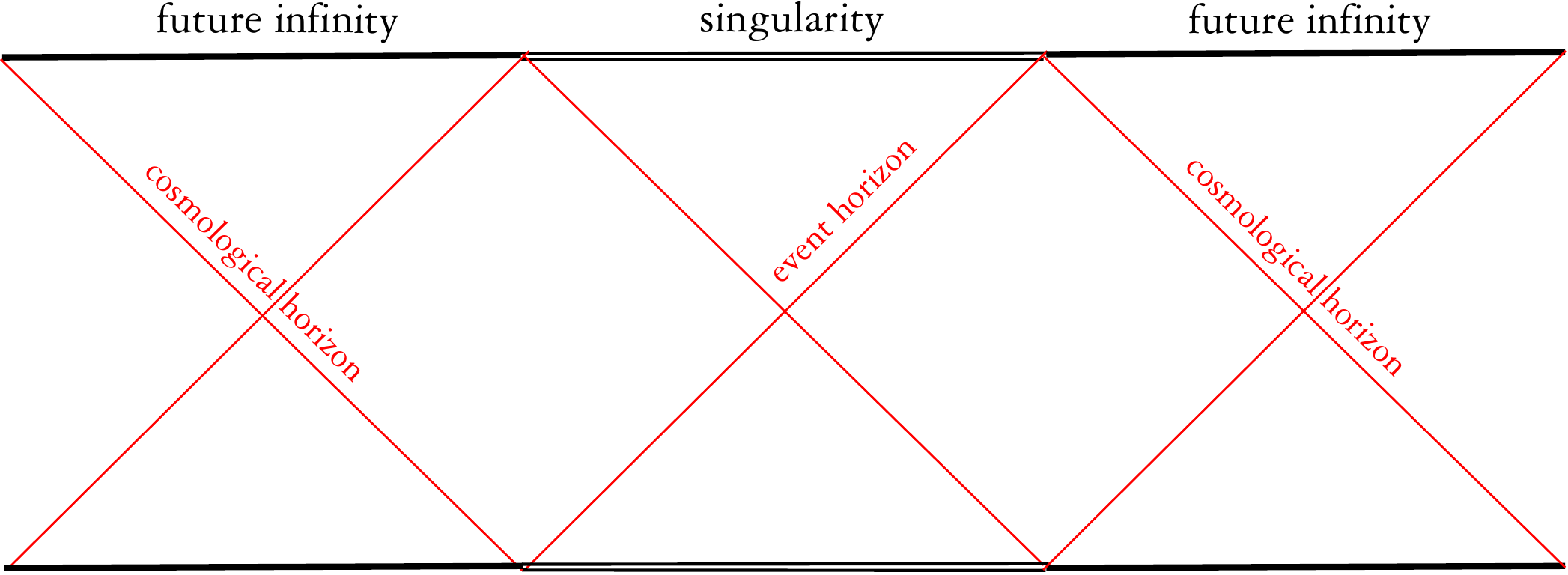} 
\caption{\textbf{Left:} The Penrose diagram of a Schwarzschild de-Sitter black hole, $\text{SdS}[\text{S}^3]$. Here the diagram extends indefinitely toward the right, as well as the left hand side.\label{penrose2}} 
\end{figure}

\begin{figure}
\includegraphics[width=2.8in]{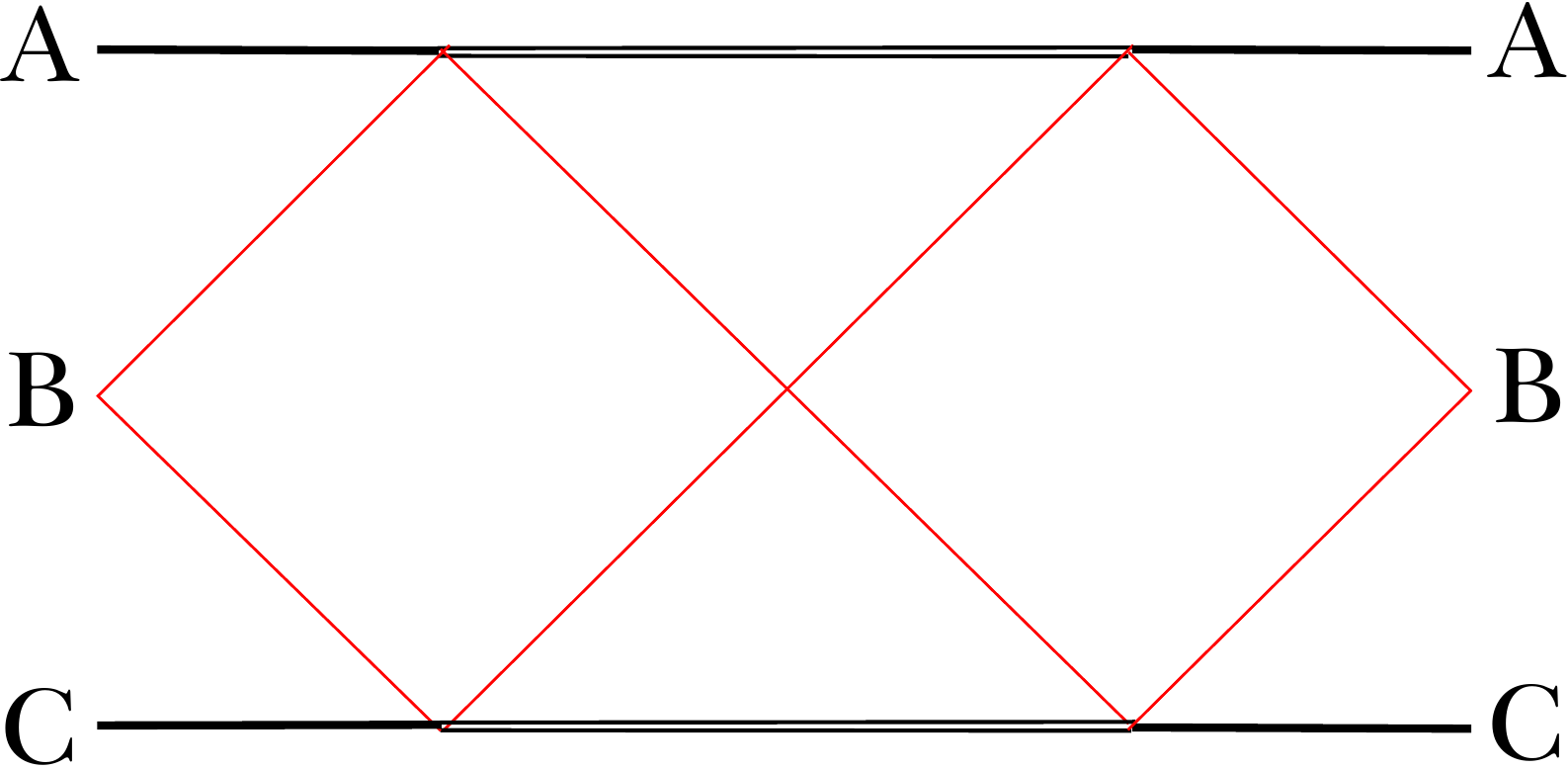} 
\caption{\textbf{Left:} The Penrose diagram of a Schwarzschild de-Sitter black hole, $\text{SdS}[\text{S}^3]$, but with topological identification, so that instead of indefinitely many black holes as in Fig.(\ref{penrose2}), we only have one black hole. \label{penrose2b}} 
\end{figure}

To our modern eyes, it does not seem that $\text{dS}[\mathbb{R}\text{P}^3]$ with a nontrivial topology is ``simpler'' than $\text{dS}[\text{S}^3]$.
To better appreciate what de Sitter meant, let us reproduce the example in \cite{0308022}, which is to consider a Schwarzschild-de Sitter black hole. To this end, let us start with a stellar collapse in $\text{dS}[\text{S}^3]$, which produces a black hole at the pole $r=0$. The resulting Schwarzschild-de Sitter metric tensor reads
\begin{flalign}
g[\text{SdS}[\text{S}^3]]=&-\left(1-\frac{2M}{r}-\frac{r^2}{\ell^2}\right)\d t^2 + \frac{\d r^2}{1-\dfrac{2M}{r}-\dfrac{r^2}{\ell^2}}\notag \\ &+ r^2(\d\theta^2 + \sin^2\theta \d\phi^2).
 \end{flalign}
Here, the notation $\text{SdS}[\text{S}^3]$ only serves to remind us that for $M=0$, we recover a de Sitter space with $\text{S}^3$ cross section.
The topology of this spacetime with a black hole is of course not $\mathbb{R} \times \text{S}^3$.
Since there are two locations that correspond to $r=0$ (namely the north and the south poles of the $\text{S}^3$), introducing a black hole at one pole apparently also implies the existence of another black hole at the other pole. In fact, this then forces us to add more black holes \emph{ad infinitum}, and the resulting Penrose diagram of Schwarzschild-de Sitter black hole extends indefinitely in the horizontal direction, as in Fig.(\ref{penrose2}). This scenario is not physical, since we do not expect that a \emph{local} gravitational collapse would produce another black hole at the ``other side'' of the universe, much less \emph{infinitely many} black holes. A possible way out of this conundrum is to topologically identify the spacetime such that one obtains what naively looks like the maximally extended version of an asymptotically flat Schwarzschild black hole (the Kruskal-Szekeres spacetime), as shown in Fig.(\ref{penrose2b}). At this point, one should ask: does this picture now represent a physically acceptable geometry? In the asymptotically flat case, the answer is yes --- in the sense that in a realistic gravitational collapse, the interior is replaced by the infalling stellar matter so that the second asymptotically flat region on the other side of the Einstein-Rosen bridge is removed. However, this is not the case here. Due to the topological identification, the second region that should be removed in a realistic gravitational collapse is, in fact, \emph{identical} to the exterior spacetime. That is, to quote \cite{0308022}, ``\emph{the `other' universe which we so casually consign to non-existence is in fact our own.}'' The conclusion is that neither the picture in Fig.(\ref{penrose2}) nor the one in  Fig.(\ref{penrose2b}) can be considered physical. However, if we take de Sitter space to be $\text{dS}[\mathbb{R}\text{P}^3]$, then this problem does not arise, since introducing a black hole at ``$r=0$'' does not entail another black hole at the other pole, simply because there is \emph{no other pole}!

Despite the aforementioned discussion, since the universe is likely to be large enough that its global topology cannot be observed directly\footnote{Nontrivial topologies are not ruled out by current observations, see, e.g., \cite{1303.5086, 1601.03884, 1604.02179}.}, one is tempted to argue that whether $\text{dS}[\mathbb{R}\text{P}^3]$ should be preferred over $\text{dS}[\text{S}^3]$ is merely based on philosophical or aesthetic grounds, or at best, on mathematical grounds. However, this is not the case --- while there is no problem with quantum field theories in $\text{dS}[\mathbb{R}\text{P}^3]$, there \emph{are} physical observables that could \emph{in principle} tell us whether our universe is indeed $\text{dS}[\mathbb{R}\text{P}^3]$, or $\text{dS}[\text{S}^3]$ \cite{9812056}. For example, an inertial observer who couples to a free scalar field through a monopole detector could distinguish the difference between these two spaces, although the difference become exponentially small in the
distant past or future on the observer's worldline \cite{9812056}. (See also \cite{0510049}.)

In \cite{0308022}, McInnes argued that the distinction between  $\text{dS}[\mathbb{R}\text{P}^3]$ and $\text{dS}[\text{S}^3]$ could be an important one in the context of \emph{quantum cosmology}. In particular, the distinction could be important in the attempt to understand the horizon entropy of a de Sitter universe --- while $\text{dS}[\mathbb{R}\text{P}^3]$, with its richer structure at conformal infinity, allows a relatively straightforward interpretation of the horizon entropy as entanglement entropy, surprisingly, $\text{dS}[\text{S}^3]$ \emph{does not}. We shall elaborate on this interesting but subtle point.

Since the horizon entropy of de Sitter space shares the same properties with the Bekenstein-Hawking entropy of black holes, i.e., it is proportional to the horizon area in Planck units \cite{GH} (it is a quarter of the horizon area in the case of $\text{dS}[\text{S}^3]$, but half of the horizon area in $\text{dS}[\Bbb{R}\text{P}^3]$, see \cite{0308022}), and it exhibits a temperature much like that of Hawking temperature, it is likely they both have the same underlying physical origin. Lately, 
it seems increasingly likely that black hole entropy is indeed entanglement entropy\footnote{By this statement, we mean the following: the state across a spatial hypersurface should -- by unitarity -- remain pure at all time if we consider the total Hilbert space. However, an exterior observer should trace over the eigenstates in the black hole subsystem and so sees a thermal state with the von Neumann entropy associated with the reduced density matrix.} (though some questions remain --- see \cite{1104.3712,9409015}). However, 
\emph{de Sitter space is quite unlike black holes}. Consider, for simplicity, an asymptotically flat two-sided eternal black hole. Its Bekenstein-Hawking entropy can be interpreted as the entanglement entropy of the two asymptotically flat regions. It is crucial that these two regions are causally separated and thus can be treated as \emph{independent} copies of the system. In the case of $\text{dS}[\text{S}^3]$, however, the spacetime region beyond the cosmological horizon, despite being causally disconnected, shares the same conformal infinities with the patch inside the horizon. Worse still, $\text{dS}[\text{S}^3]$ does not actually have a wormhole that connects two independent asymptotically regions, it has instead a ``handle'' that loops back to the \emph{same} universe \cite{0308022}. Mathematically, the entanglement entropy interpretation for asymptotically flat two-sided black holes work because the connected sum of two copies of $\Bbb{R}^3$ is nontrivial: $\Bbb{R}^3 \# \Bbb{R}^3 \neq \Bbb{R}^3$, a necessary condition for the two sides to remain independent. For  $\text{dS}[\text{S}^3]$ the scheme runs into problem because $\text{S}^3 \# \text{S}^3 = \text{S}^3$. However, fortunately, $\mathbb{R}\text{P}^3 \# \mathbb{R}\text{P}^3 \neq \mathbb{R}\text{P}^3$. (For more details, see \cite{0308022}; here we only reproduce the gist of the argument for completeness.)
So it would seem that the $\text{dS}[\mathbb{R}\text{P}^3]$ picture does have some advantages. Moreover, as we have mentioned, black holes in $\text{dS}[\mathbb{R}\text{P}^3]$ have causal structures that can be argued to be more physical than that of a black hole in $\text{dS}[\text{S}^3]$. (A black hole in $\text{dS}[\mathbb{R}\text{P}^3]$ has some similarities with the so-called ``geon black holes'' \cite{9906031v1, 1001.0124v1}, but they are different geometries altogether. Recently, the idea of topologically identifying the antipodal points on a black hole horizon was also proposed \cite{1601.03447} in an attempt to resolve the information paradox, but this identification is also different from the one we are considering here.)

In view of the importance to understand $\text{dS}[\mathbb{R}\text{P}^3]$, especially in the context of quantum gravity, in this work we investigate the instanton tunneling of $\text{dS}[\mathbb{R}\text{P}^3]$, and show that the global topology of the spatial sections of de Sitter space can indeed influence the tunneling physics. 

\section{Was there Something or Nothing?}

A major research program in quantum cosmology concerns the study of the wave function of the universe, known as the Wheeler-DeWitt equation \cite{DeWitt}. The ground state of the wave function can be derived from a Euclidean path integral, which is schematically of the form
\begin{equation} 
{\Psi[h_{\mu\nu}^{\text{out}},\Phi^\text{out};h_{\mu\nu}^{\text{in}},\Phi^\text{in}]}= \int \mathcal{D}[g]\mathcal{D}[\Psi] e^{-I[g_{\mu\nu},\mathcal{F}]},
\end{equation}
involving a sum over all possible geometries, and over all histories that are connected by two hypersurfaces. 
Here $\Phi$ is the restriction of the field $\mathcal{F}$ -- which represents matter degrees of freedom -- on a spacelike hypersurface with Riemannian metric $h_{\mu\nu}$.
(If the Euclidean geometry is not connected, i.e., in the ``nothing-to-something'' scenario -- see below -- we remove the initial hypersurface $h^\text{in}$ and $\Phi^\text{in}$.)
The Euclidean path integral is a good approximation of the ground state wave function, which can be approximated by solving for on-shell solutions known as ``instantons''. 
Specifically the wave function is to a good approximation,  $\Psi \propto\exp(-I_{\text{instanton}})$, where $I_{\text{instaton}}$ is the action of a saddle-point solution which satisfies the Euclidean Einstein's field equations with some prescribed boundary conditions.
Instantons provide a mean to calculate the probability of one geometry tunneling into another, $\text{Prob} =\Psi^*\Psi\propto \exp[-2\cdot\text{Re}(I_{\text{instanton}})]$.

One possibility is that the universe was initially some Lorentzian spacetime, but underwent a tunneling into a different Lorentzian spacetime. Such a picture is provided by, e.g., the Farhi-Guth-Guven tunneling \cite{FGG}, and Fischler-Morgan-Polchinski tunneling \cite{FMP1, FMP2}. Mathematically, an instanton is a complete non-singular Euclidean manifold that ``connects'' these two Lorentzian spacetimes. In the language of \cite{1512.03914}, such kind of tunneling is dubbed ``something-to-something''. 
(Of course, it is possible that not everywhere in the initial spacetime undergoes such tunneling, but only a ``small'' patch, which leads to a ``pocket universe''.)
However, an alternative picture exists. It could be that the universe was initially Euclidean (by which we really mean that the geometry is Riemannian), but underwent a ``tunneling'' from ``nothingness'' to a Lorentzian spacetime. (``Nothing'' here simply refers to a lack of a classical spacetime.) One example of such a scenario is the ``No Boundary proposal'' of Hartle and Hawking \cite{HH}. This corresponds to the ``nothing-to-something'' scenario of \cite{1512.03914}.

To be more specific, ``something-to-something'' means that the initial hypersurface $\Sigma_0$ of a Lorentzian manifold $\mathcal{M}_0$ is connected to the final hypersurface $\Sigma_1$ of another Lorentzian manifold $\mathcal{M}_1$ via a Euclidean manifold $\mathcal{M}_\text{E}$. More precisely, by ``connected'', we mean the following: start with $\mathcal{M}_0$ foliated by a family of spacelike hypersurfaces $\Sigma_t$, where $t \in [-a,0]$ corresponds to the time direction in which $\Sigma_t$ evolves via Einstein's equations. At time $t=0$, we perform an analytic continuation (a Wick rotation) such that the metric signature is now $(+,+,+,+)$. We can perform another analytic continuation to return to the Lorentzian signature, but the second analytic continuation need not bring the geometry back to its initial configuration, instead it could be a different spacetime altogether, $\mathcal{M}_1$, which is now foliated by $\Sigma_s$, where $s \in [0,b] $, where $s$ is a time coordinate not necessarily the same as $t$. A new spacetime is thus created at the moment $s=0$.
On the other hand, in the ``nothing-to-something'' scenario, $\Sigma_0$ and $\Sigma_1$ are disconnected. 

A mathematically rigorous way to understand Wick-rotating spacetimes, is to consider a complex manifold $\mathcal{M}_\Bbb{C}$, with real Lorentzian section $\mathcal{M}_L$ and a real Riemannian section $\mathcal{M}_R$, with non-empty intersection $\Sigma$ where one geometry can ``rotate'' to another via (anti)holomorphic involutions  --- more on this later. 
$\mathcal{M}_R$ is usually taken to be orientable, connected, and compact with boundary $\Sigma$.
An instanton, or more specifically, a gravitational instanton, is simply the ``doubling'' of  $\mathcal{M}_R$. That is, if we take two copies of  $\mathcal{M}_R$, and call them  $\mathcal{M}_R^+$ and  $\mathcal{M}_R^-$ respectively, then an instanton is simply  $2\mathcal{M}_R=\mathcal{M}_R^+ \cup  \mathcal{M}_R^-$, joined over $\Sigma$. 

In the following discussion, we shall illustrate both the ``something-to-something'' and ``nothing-to-something'' scenarios using $\text{dS}[\text{S}^3]$ as a concrete example. 
A Euclidean de Sitter space, with Euclidean time $\eta$, can be described by the metric of the form
\begin{equation}\label{EdS}
g[\text{EdS}] = \d\eta^2 + \rho^2(\eta) [\d\chi^2 + \sin^2\chi (\d\theta^2 + \sin^2 \theta \d\phi^2)],
\end{equation}
where $\rho=\ell \sin(\eta/\ell)$, with $\ell$ being the length scale of de Sitter space associated with the cosmological constant, $\Lambda=3/\ell^2$.
The coordinate range for $\eta$ and $\chi$ are, respectively, $0 \leqslant \eta/\ell\leqslant \pi$, and $0 \leqslant \chi \leqslant \pi$. 
By performing a Wick rotation $\eta = \pi \ell/2 + it$, we see that $\d\eta^2 = -\d t^2$, and
\begin{equation}\label{etarot}
\sin\left(\frac{\eta}{\ell}\right)=\sin\left(\frac{\pi}{2} + \frac{it}{\ell}\right) = \cosh\left(\frac{t}{\ell}\right).
\end{equation}
We therefore recover the metric $g[\text{dS}]=-\d t^2 + g[\text{S}^3]$, where $g[\text{S}^3]$ is given by Eq.(\ref{S3}).
The Lorentzian version of this geometry is clearly time dependent: it describes the well-known de Sitter hyperboloid, with a family of 3-spheres that grows indefinitely in size toward the future as well as ``toward'' the past. In fact it covers the entire spacetime, and is known as the global coordinates. The topology of de Sitter space as $\mathbb{R} \times \text{S}^3$ is evident from this metric.
In the language of complex manifold, we can consider $\mathcal{M}_\Bbb{C}$ as a complex quadric in $\Bbb{C}^5$, which would contain as real sections both de Sitter space $\mathcal{M}_L=\text{dS}[\text{S}^3]$, and the sphere $\mathcal{M}_R=\text{S}^4$, such that $\mathcal{M}_L \cap \mathcal{M}_R = \text{S}^3$. See, e.g., \cite{1110.0611v1}, for further discussion. 
However, this is not the only way to obtain a Lorentzian metric from $g[\text{EdS}] $, as we shall discuss below.

In addition to the global coordinates, there exists also static coordinates in which the metric is time-independent. The Euclidean version of that, with Euclidean time $\tau$, is 
\begin{flalign}
g[\text{Static EdS}] =& \left(1-\frac{r^2}{\ell^2}\right)\d\tau^2 + \left(1-\frac{r^2}{\ell^2}\right)^{-1}\d r^2 \notag \\ &+ r^2(\d\theta^2 + \sin^2 \theta \d\phi^2),
\end{flalign}
where $0 \leqslant r \leqslant \ell$, and $-\pi/\ell\leqslant \tau/\ell\leqslant \pi/\ell$.
Of course in the Euclidean case, there is no time. It is to be understood that by ``static'' coordinates we meant that the Lorentzian version of the geometry, obtained by Wick-rotating $\tau$, is static.

The two metrics are related via the transformation rules:
\begin{equation}
r = \ell\sin\left(\frac{\eta}{\ell}\right) \sin \chi; ~~~ \tan\left(\frac{\tau}{\ell}\right) = \tan\left(\frac{\eta}{\ell}\right) \cos \chi.
\end{equation}

The global coordinates can be seen to represent a 4-dimensional sphere, $\text{S}^4 \hookrightarrow \Bbb{R}^5$, with the canonical flat metric
\begin{equation}
g[\mathbb{R}^5]=\eta_{ij}\d x^i\d x^j, ~~~x^i \in \left\{1,2,3,4,5\right\}, 
\end{equation}
where
\begin{equation}\notag
\begin{cases}

x^1 =\ell \cos (\eta/\ell),\\

x^2 = \ell \sin (\eta/\ell) \cos \chi, \\

x^3 = \ell \sin(\eta/\ell) \sin \chi \cos \theta, \\

x^4 = \ell\sin(\eta/\ell) \sin \chi \sin\theta \cos \phi, \\

x^5 =\ell\sin(\eta/\ell) \sin \chi  \sin\theta\sin\phi. 

\end{cases}
\end{equation}

\begin{figure}
\includegraphics[width=2.8in]{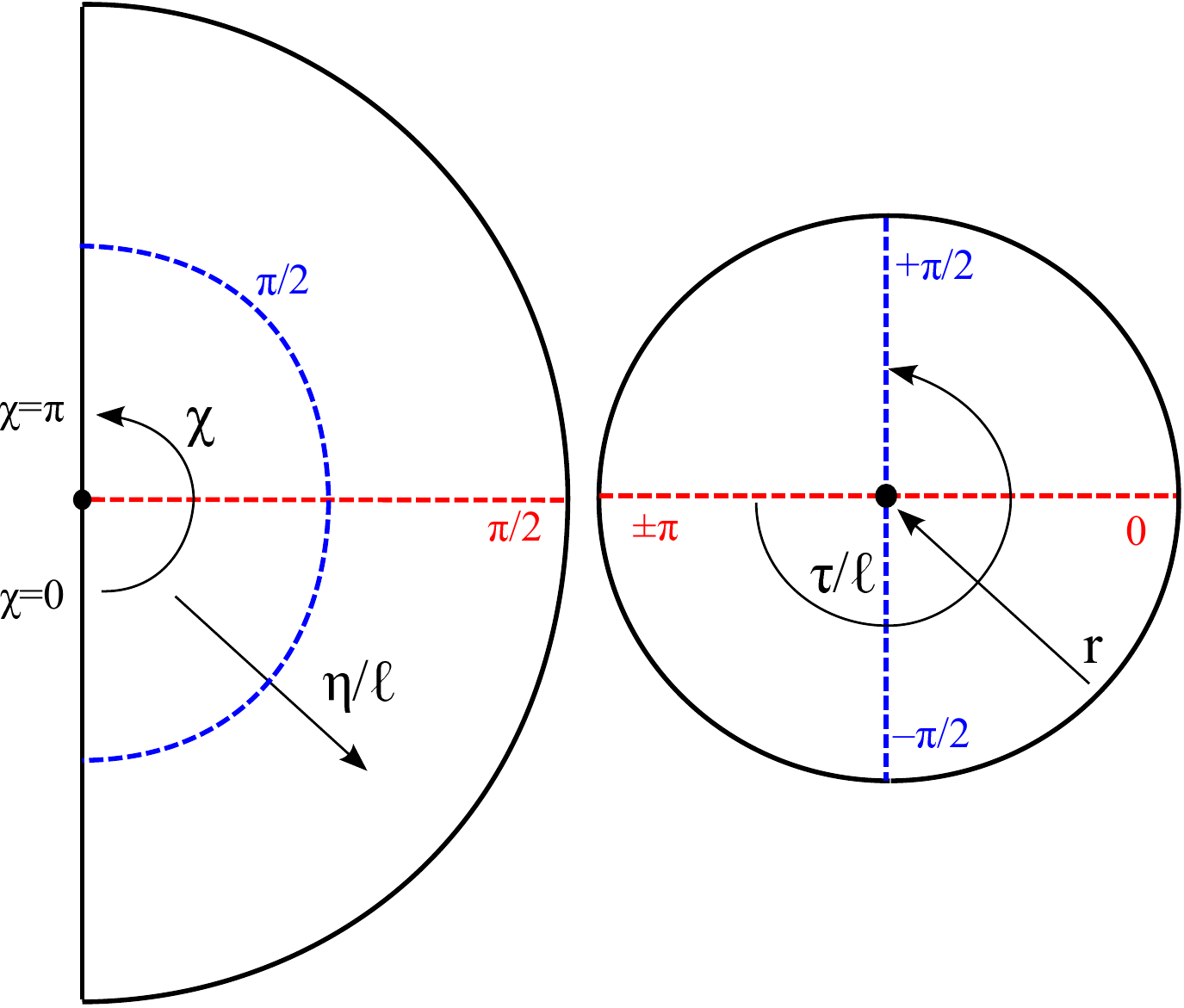} 
\caption{The Euclidean version of $\text{dS}[\text{S}^3]$, in the $\eta$-$\chi$ global coordinates (left) and the $\tau$-$r$ static coordinates (right). 
See the main text for a detailed explanation of these coordinates.
To obtain a homogeneous tunneling, a Wick rotation is performed along the blue curves; to obtain an inhomogeneous tunneling, a Wick rotation is performed along the red curves. \label{desitter}} 
\end{figure}

We see that the coordinates $(\eta/\ell,\chi,\theta,\phi)$ are angular coordinates on the $\text{S}^4$ with radius $\ell$. 
We have $0\leqslant  \eta/\ell,\chi,\theta \leqslant \pi$ and $0 \leqslant \phi < 2\pi$. This is just a straightforward generalization of the usual spherical coordinates.
In Fig.(\ref{desitter}), the left hand figure represents part of this coordinate system. Note that unlike the 2-sphere, whose angular coordinates are $0\leqslant \vartheta \leqslant \pi$ and $0 \leqslant \varphi <2\pi$, respectively, here both $\eta/\ell$ and $\chi$ range from 0 to $\pi$. 

Similarly, the static coordinate can be seen to be representing a $\text{S}^4  \hookrightarrow \Bbb{R}^5$ with the canonical flat metric 
\begin{equation}
g[\mathbb{R}^5]=\eta_{ij}\d y^i\d y^j, ~~~y^i \in \left\{1,2,3,4,5\right\}, 
\end{equation}
where
\begin{equation}\notag
\begin{cases}

y^1 = r \sin \theta \cos \phi, \\

y^2 = r \sin \theta \sin \phi, \\

y^3 = r \cos \theta, \\

y^4 = \sqrt{\ell^2 - r^2} \cos(\tau/\ell), \\

y^5 = \sqrt{\ell^2-r^2} \sin(\tau/\ell). 

\end{cases}
\end{equation}

\begin{figure}
\includegraphics[width=3.52in]{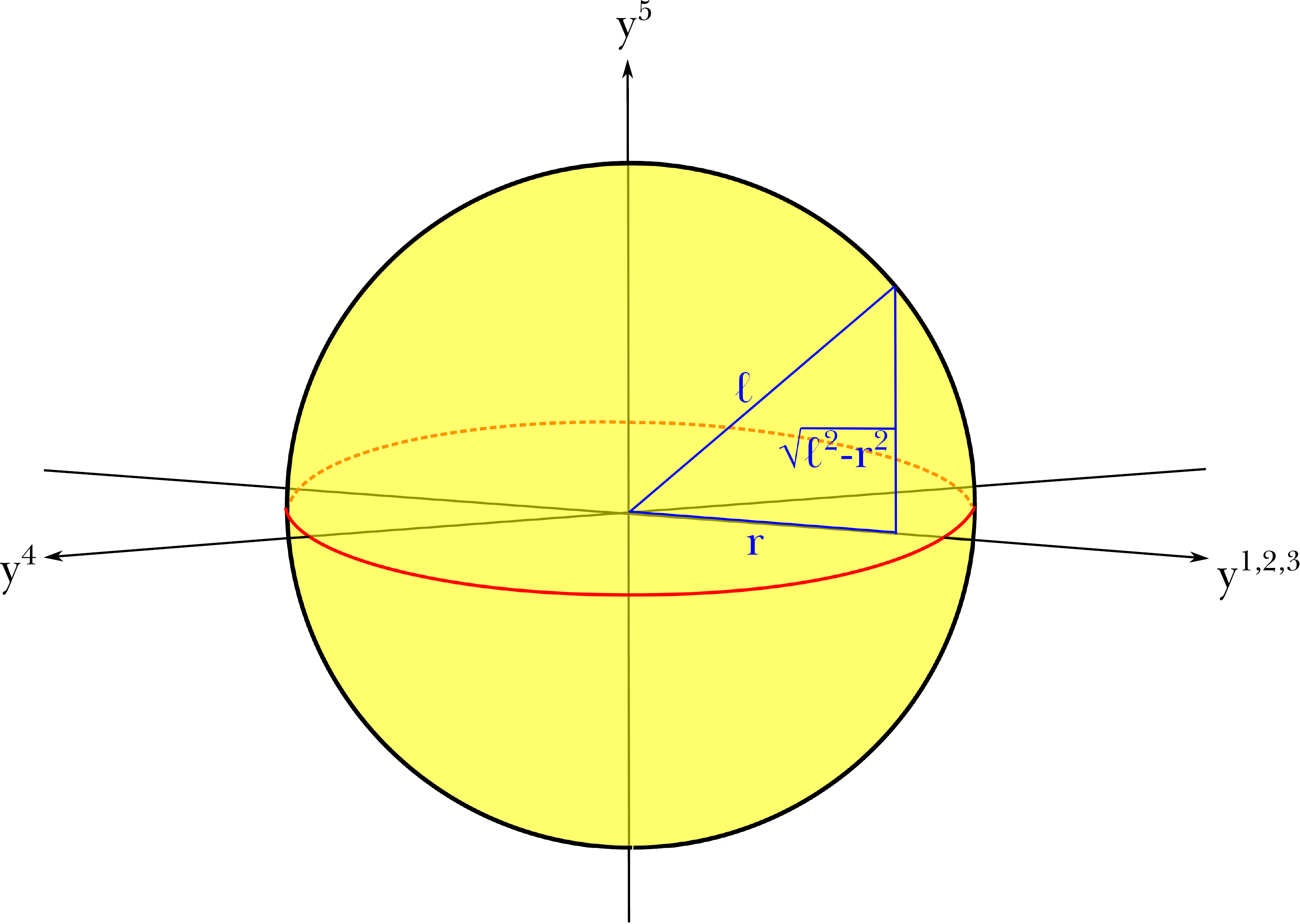} 
\caption{\textbf{Left:} The 4-dimensional sphere shown parametrized by the coordinate $r$. The horizontal slice in red corresponds to $\tau/\ell=\pi$. \label{3dsphere}} 
\end{figure}

In Fig.(\ref{desitter}), the right hand figure represents part of this coordinate chart. It can be thought as the projection of the $\text{S}^4$ onto the $y^4$$y^5$-plane. 
Note that $r=0$ corresponds to the circle $(y^4)^2 + (y^5)^2=\ell^2$, with $y^1=y^2=y^3=0$. This is the boundary circle of the right hand figure of Fig.(\ref{desitter}).
Likewise, $r=\ell$ corresponds to the center $y^4=y^5=0$.
In Fig.(\ref{3dsphere}), we show how the coordinate $r$ parametrizes the $\text{S}^4$. 
The horizontal slice in red corresponds to $\tau/\ell=\pi$, that is, it corresponds to the red dotted line on the right diagram of Fig.(\ref{desitter}).

There are in fact two kinds of tunnelings one can consider --- a homogeneous one and an inhomogeneous one. If, after an analytic continuation to Lorentzian signature one finds that a bubble is formed, then the spacetime becomes inhomogeneous, and we refer to the tunneling as inhomogeneous. An example of this is the Coleman-De Luccia tunneling \cite{CDL}.  On the other hand, if the field value of the entire universe tunnels from one value to another, then the tunneling is homogeneous. An example would be the Hawking-Moss tunneling \cite{HM}. 
Homogeneous tunneling for $\text{dS}[\text{S}^3]$ can be achieved by performing Wick rotation along the $\eta/\ell = \pi/2$ slice in the global coordinates, as we have done before in the discussion around Eq.(\ref{etarot}). This is shown as the blue dashed semi-circle on the left of Fig.(\ref{desitter}). This is equivalent to the blue dash line in the static coordinates, depicted on the right of Fig.(\ref{desitter}). 
On the other hand, to achieve an inhomogeneous tunneling, we should instead perform a Wick rotation along the $\chi=\pi/2$ slice in the global coordinates, shown as the red dashed line on the left of Fig.(\ref{desitter}). In the static coordinate, this corresponds to the red dashed line shown on the right of Fig.(\ref{desitter}). Note that at the level of pure geometry, whether we Wick-rotate along the blue line or the red line makes no difference due to spherical symmetry; the difference only arises when one considers the form of the \emph{potential} (of, usually, a scalar field) that governs the tunneling. When one does consider an appropriate potential though, the inhomogeneous case has several advantages. For example, it allows one to consider an open, $k=-1$, cosmology, whereas the homogeneous case is restricted to the closed, $k=1$, cosmology \cite{9802030}.

\begin{figure}
\includegraphics[width=3.5in]{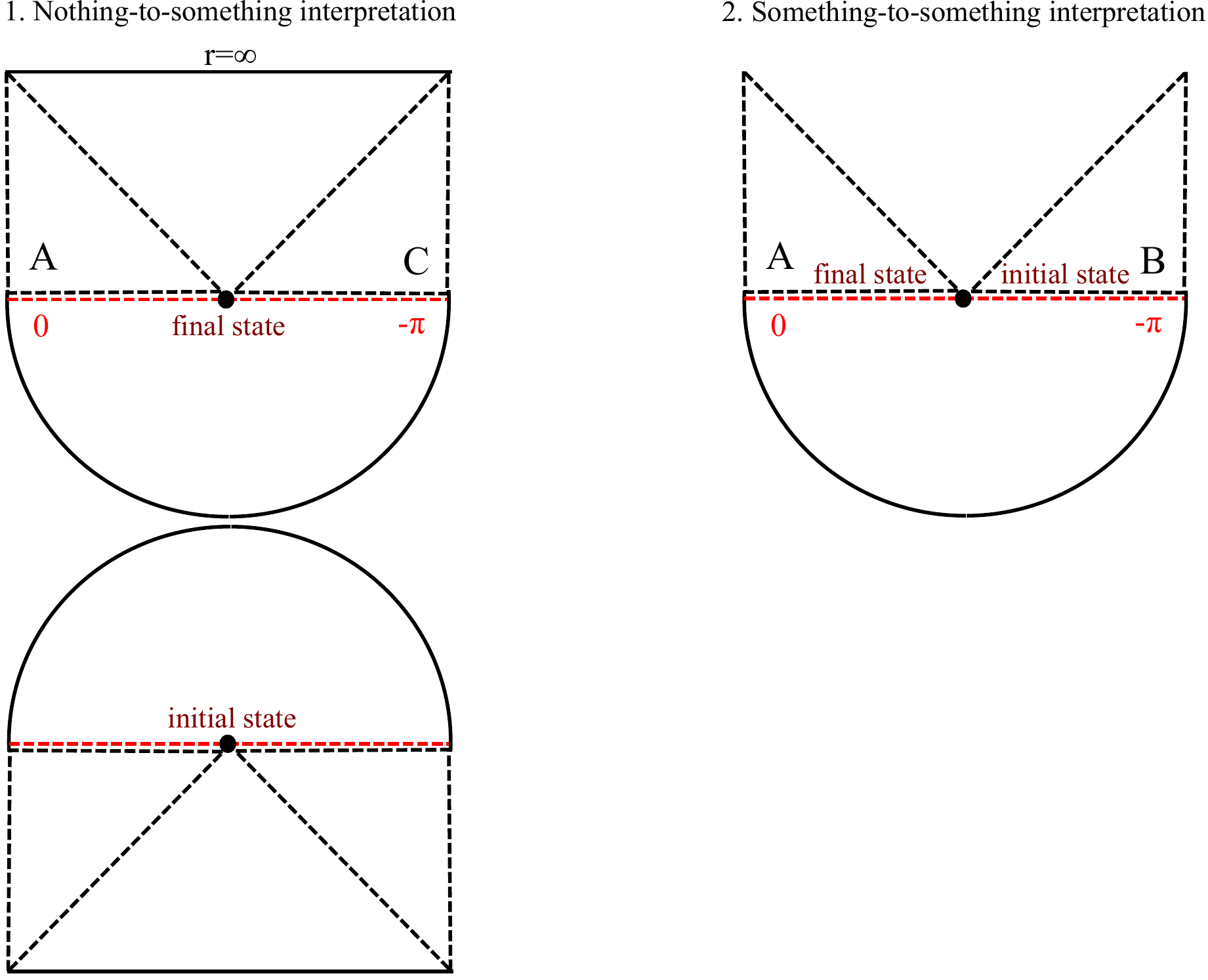} 
\caption{\textbf{Left:} The ``nothing-to-nothing'' interpretation; and \textbf{Right:} the ``something-to-something'' interpretation.\label{interpretation}} 
\end{figure}

As already explained in \cite{1512.03914}, there are two possible interpretations for the inhomogeneous case. The easiest way to perform the Wick-rotation to Lorentzian signature, is to simply glue the entire hypersurface of the Euclidean manifold onto the Lorentzian one, as shown on the left of Fig.(\ref{interpretation}). The interpretation is that a Lorentzian de Sitter space $\text{dS}[\text{S}^3]$ is created out of nothing. This is the ``nothing-to-something'' interpretation. The initial state and the final state include both of the Lorentzian causal patches, but the initial state is not connected to the final state --- the instanton corresponds to two
disconnected halves of $\text{S}^4$.  Physically, an (eternal) de Sitter space $\text{dS}[\text{S}^3]$ fluctuates into nothing and back again. In a more realistic cosmology, we do not usually consider the contracting phase. In this case, we take only the top half of the left figure in Fig.(\ref{interpretation}), and interpret the tunneling as ``creation from nothing'' \`a la Hartle and Hawking \cite{HH}.

An alternative interpretation of the tunneling was proposed by Brown and Weinberg \cite{0706.1573}, in which the $t=0$ hypersurface of the right causal patch of the Lorentzian de Sitter space is treated as the ``initial state'', while the $t=0$ hypersurface of the left causal patch is treated as the ``final state''. (Of course the choice of left and right here is completely arbitrary.) The initial state is ``rotated'' through the Euclidean manifold to match with the final state. See the right figure of Fig.(\ref{interpretation}). This is the ``something-to-something'' interpretation. While this scenario is not as mathematically natural as the ``nothing-to-something'' interpretation, it does have one advantage: since the initial state and final state is connected via the Euclidean manifold, the left and right causal patches are ``naturally'' related; whereas in the case of ``nothing-to-something'' scenario, the tunneling needs to cover both causal patches, one of which is not accessible to the observer in the other patch (and thus, is arguably not physically relevant). 

Such an argument that involves causal patches of de Sitter space should now be familiar to us --- it is exactly the sort of issues that are relevant to the debate whether dS[$\mathbb{R}\text{P}^3$] should be preferred over dS[$\text{S}^3$]. In the case of dS[$\mathbb{R}\text{P}^3$], there is \emph{only one} causal patch. One thus naturally wonders, whether in the case of dS[$\mathbb{R}\text{P}^3$], the instanton tunneling still admits both ``nothing-to-something'' and ``something-to-something'' interpretations. However, the first question one faces when trying to understand tunneling physics for dS[$\mathbb{R}\text{P}^3$] is the following: \emph{Does this geometry admit a well-defined Euclidean counterpart?} The answer is yes, but it is not straightforward to describe the corresponding geometry.

\section{Instanton Tunneling for\newline de Sitter Space with $\mathbb{R}\text{P}^3$ Topology}

Recall that the antipodal map on $\text{S}^3$ that yields a $\mathbb{R}\text{P}^3$  is given by
\begin{equation}
\mathcal{A}: \text{S}^3 \to \text{S}^3
\end{equation}
such that $\chi \mapsto \pi - \chi$, $\theta \mapsto \pi-\theta$, and $\phi \mapsto \pi +\phi$. 
In order to obtain dS[$\mathbb{R}\text{P}^3$], we can start with $\text{S}^4$ with an appropriate topological identifications followed by a proper Wick-rotation. One might be tempted to start with an $\text{S}^4$, and make the topological identifications on the angular coordinates following the antipodal map $\mathcal{A}$. Note that this leaves one angular coordinate of $\text{S}^4$ untouched. So the Euclidean geometry is a real projective space immersed in a 4-sphere: $\mathbb{R}\text{P}^3 \looparrowright \text{S}^4 \hookrightarrow \mathbb{R}^5$. Note that the manifold $\mathbb{R}\text{P}^3$ cannot be embedded in $\text{S}^4$. In fact this is true in general dimensions: $\mathbb{R}\text{P}^{n-1}$ cannot be embedded in $\text{S}^{n}$, a theorem in cohomology theory which can be shown using a Mayer-Vietoris type argument. This geometry is therefore not a very ``nice'' one. Furthermore, such a topological identification is \emph{not} what we want for the following reason: if we were to perform a Wick-rotation at $\chi = \pi/2$ hypersurface, then we would ``unwind'' $\chi$ into a temporal coordinate. Among the remaining 3 spatial coordinates, only two (namely $\theta$ and $\phi$) are acted upon by $\mathcal{A}$. So the spatial topology is \emph{not} $\mathbb{R}\text{P}^3$, and thus we cannot obtain in this way dS[$\mathbb{R}\text{P}^3$]. 

One might be tempted to think that it is better to consider the Euclidean manifold to be $\mathbb{R}\text{P}^4$, which is obtained from the antipodal map
\begin{equation}
\tilde{\mathcal{A}}: \text{S}^4 \to \text{S}^4,
\end{equation}
such that $\psi \mapsto \pi - \psi$, $\chi \mapsto \pi - \chi$, $\theta \mapsto \pi-\theta$, and $\phi \mapsto \pi +\phi$,
where $\left\{\psi, \chi, \theta, \phi\right\}$ are the angular coordinates on $\text{S}^4$. Unlike $\mathbb{R}\text{P}^3 \looparrowright \text{S}^4$, the manifold $\mathbb{R}\text{P}^4$ is a ``well-behaved'' smooth manifold, and most importantly, upon Wick-rotating $\chi$, one would obtain a Lorentzian spacetime with  $\mathbb{R}\text{P}^3$ as spatial sections. It is true that $\mathbb{R}\text{P}^4$ is not orientable, unlike $\mathbb{R}\text{P}^3$, but this is not a concern here since $\mathbb{R}\text{P}^4$ is the Euclidean geometry, and there is no requirement that the Euclidean manifold should be orientable. (On the other hand, the spatial slices of a Lorentzian manifold must be orientable, otherwise this would contradict the observation that parity is violated in our universe (see for example ref.(\cite{visser}), page 289). Non-orientable instanton has indeed been studied in the literature \cite{9607079v2}. (Recently, non-orientable manifolds also received some attentions in holography \cite{1603.04426}.) However, it turns out that $\mathbb{R}\text{P}^4$ is \emph{not} a mathematically viable option \cite{Ratcliffe}. This is due to a topological constraint: the Euler characteristic $\chi$ of a 4-dimensional instanton (not to be confused with the coordinate $\chi$) must be even:
\begin{flalign}
\chi(2\mathcal{M}_R)&=\chi(\mathcal{M}_R^+ \cup \mathcal{M}_R^-) \\
&= \chi(\mathcal{M}_R^+) + \chi(\mathcal{M}_R^-) - \chi(\Sigma) \\
&= 2\chi(\mathcal{M}_R^+).
\end{flalign}
Here we have made used of the fact that $\chi(\mathcal{M}_R^+)=\chi(\mathcal{M}_R^-)$, and that $\chi(\Sigma)=0$. The latter is a well-known fact that the Euler characteristic of any closed odd-dimensional manifold is zero. The Euler characteristic of $\mathbb{R}\text{P}^4$ is however, odd: $\chi(\mathbb{R}\text{P}^4)=1$. 
Note that, while $\mathbb{R}\text{P}^4$ does not fit in the framework discussed here, one can nevertheless consider $\mathbb{R}\text{P}^4$ in other contexts. See, e.g., \cite{9805101}, in which $\mathbb{R}\text{P}^4 \setminus \text{B}^4$ was considered (where $\text{B}^4$ is the open 4-dimensional ball). However, even then, $\mathbb{R}\text{P}^4$ gives rise to time non-orientable Lorentz signature section. This time non-orientability gives rise to some problems (or at least some restrictions) in the quantum field theory defined on such a spacetime \cite{9505035}.

However, as shown in \cite{9607079v2}, there are numerous Wick-rotating constructions one could perform to obtain a spacetime, not all of which are guaranteed to be orientable, or admitting a \emph{pinor} structure that is crucial for the existence of fermions. Fortunately, from the viewpoint of complex geometry, just like its $\text{dS}[\text{S}^3]$ cousin, $\text{dS}[\mathbb{R}\text{P}^3]$ can also be regarded as a real Lorentzian section of a complex manifold, and there is no problem defining its Euclidean counterpart as the real Riemannian section, which is a quotient geometry of $\text{S}^4$ under some $\Bbb{Z}_2$ action \cite{9607079v2,9812056}. The details will not be important for our purpose, but it is worth emphasizing that the Euclidean version of $\text{dS}[\mathbb{R}\text{P}^3]$, which we shall denote $\text{EdS}[\mathbb{R}\text{P}^3]$, is not a manifold, but a $\mathbb{Z}_2$-orbifold. The reason is simple: take $\text{dS}[\text{S}^3]$ as a hyperboloid in 5-dimensional Minkowski space $\mathbb{R}^{4,1}$, one can construct $\text{dS}[\mathbb{R}\text{P}^3]$ by antipodal map on each $\mathbb{S}^3$ cross section. It is easy to see that this map has no fixed point, and so the resulting quotient is indeed a manifold. On the other hand, a similar mapping on $\text{S}^4\subset \mathbb{R}^5$ has two fixed points: the (arbitrary) north pole and south pole of the $\text{S}^4$. Due to these fixed points, the quotient geometry is not a manifold. 

In the following discussions, let us focus on the static coordinate chart, and consider some possible tunneling scenarios. However,  we will first present a \emph{non-viable} case, to illustrate how topology could rule out some scenarios.

\begin{subsection}{Case I: Something-to-Something with Euclidean Projective Space Instanton}

In the left diagram of Fig.(\ref{estatic}), we represent the topological identification of $\text{S}^4$ into $\text{EdS}[\mathbb{R}\text{P}^3]$. After the antipodal map, we need only consider half of the hemisphere, say, the bottom half of Fig.(\ref{3dsphere}), which is represented by the right diagram of Fig.(\ref{estatic}). 

\begin{figure}
\includegraphics[width=3.3in]{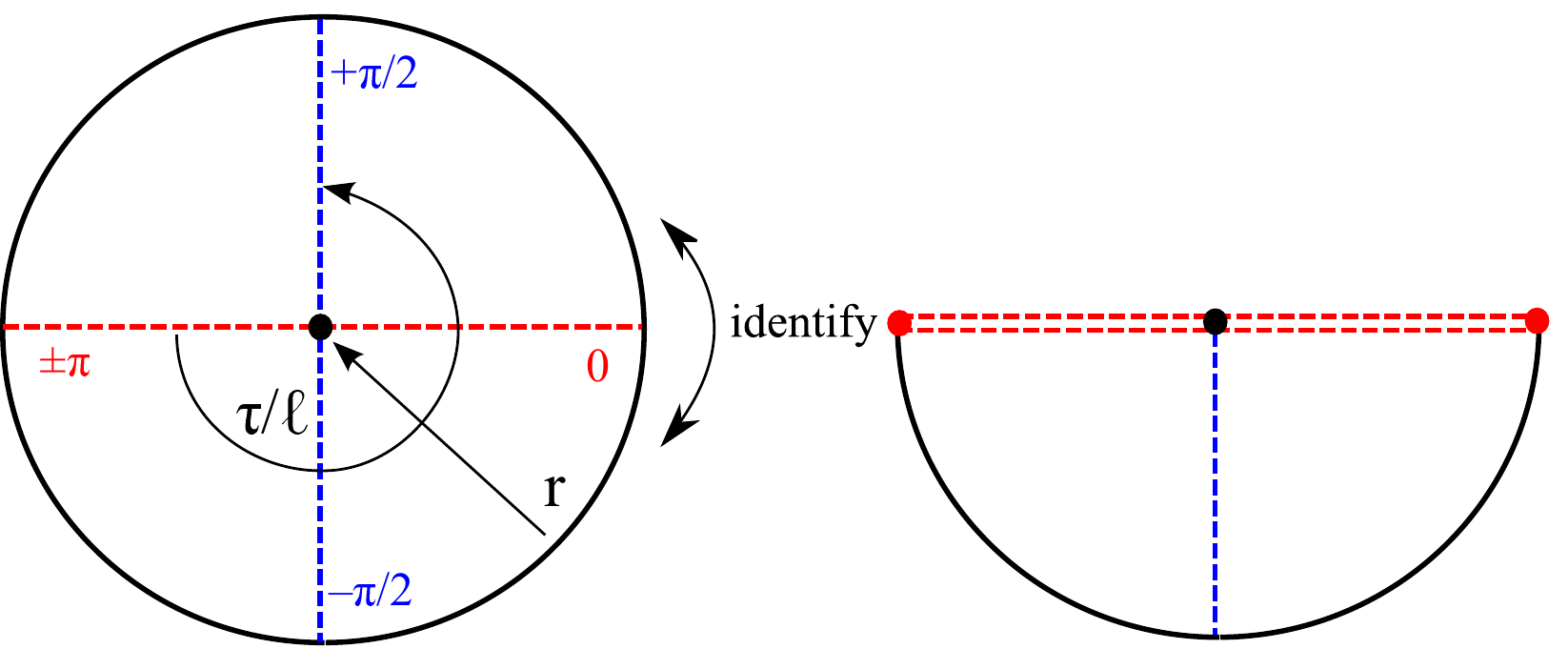} 
\caption{\textbf{Left:} The antipodal map acting on the 4-sphere; here only one of the angular coordinate $\tau/\ell$ is shown. Each vertical slice corresponds to a $\mathbb{R}\text{P}^3$. \textbf{Right:} The resulting quotient geometry consists of a hemisphere of $\text{S}^4$, each point on the boundary in red corresponds to a $\mathbb{R}\text{P}^2$. The two red dots represent the fixed points under the antipodal map. \label{estatic}} 
\end{figure}  

This naively admits a ``something-to-something'' tunneling to $\text{dS}[\mathbb{R}\text{P}^3]$. In Fig.(\ref{penrose1}), we show the Penrose diagrams of both $\text{dS}[\text{S}^3]$ and $\text{dS}[\mathbb{R}\text{P}^3]$, but with the spacelike slice $t=0$ highlighted. The significance of this slice will become clear in a moment. Note that under the antipodal map, e.g., region C in the left diagram is identified with A, so in the right diagram, we only have half of the ``original'' diagram, and again the double line on the right boundary of $\text{dS}[\mathbb{R}\text{P}^3]$ denotes the $\mathbb{R}\text{P}^2$ equator. 

%\begin{figure}
%\includegraphics[width=3.3in]{bounce_lorentz} 
%\caption{The Penrose diagram of $\text{dS}[\text{S}^3]$ (left), and $\text{dS}[\mathbb{R}\text{P}^3]$ (right). The double line on the right boundary of $\text{dS}[\mathbb{R}\text{P}^3]$ is its equator, each point of which corresponds to a $\mathbb{R}\text{P}^2$. \label{bounce_lorentz}}  
%\end{figure}  

A ``something-to-something'' tunneling can naively be constructed by pasting region A and region B of the Lorentzian $\text{dS}[\mathbb{R}\text{P}^3]$ to the Euclidean $\text{EdS}[\mathbb{R}\text{P}^3]$, so that the $t=0$ slice of region B serves as the initial state, which ``rotates'' through the Euclidean sector and ends with $t=0$ slice of section A; see Fig.(\ref{wickrotateRP3}).
Note that every point on the red double line in Fig.(\ref{wickrotateRP3}) is a $\mathbb{R}\text{P}^2$. In this picture, we take $\tau/\ell=0,\pm \pi$ to be the poles of our $\text{S}^4$, so that under antipodal map, the resulting $\mathbb{Z}_2$-orbifold has fixed points there. Due to the fixed points being exactly on the hypersurface on which the Lorentzian geometry is pasted with the Euclidean one, an obstruction arises from the nature of the fixed point. As we have mentioned, the fixed point under the antipodal map means that the geometry is not a Riemannian manifold at the neighborhood of the fixed point, but the Lorentzian spacetime is a manifold everywhere. However, more crucially, on the purported matching surface, each point on the Euclidean side is $\mathbb{R}\text{P}^2$, but the Lorentzian geometry is $\text{S}^2$ (except at the black dot, i.e., on the cosmological horizon, where it is also $\mathbb{R}\text{P}^2$). Therefore this ``matching surface'' is not allowed topologically. Such an obstruction would mean that Case I is \emph{not} an acceptable picture.

 \begin{figure}
\includegraphics[width=1.8in]{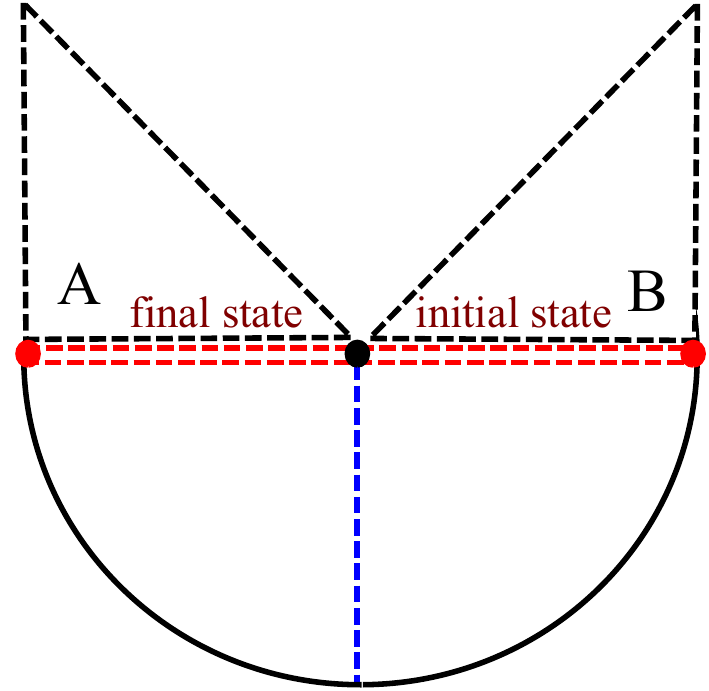} 
\caption{A purported ``something-to-something'' tunneling between $\text{EdS}[\mathbb{R}\text{P}^3]$ and $\text{dS}[\mathbb{R}\text{P}^3]$. This scenario is not possible. \label{wickrotateRP3}} 
\end{figure}  

\end{subsection}

\begin{subsection}{Case II: Nothing-to-Something with Euclidean Projective Space Instanton}

Another possibility is to consider Wick-rotating along the blue dashed line as shown in Fig.(\ref{fixedpoints}).
This gives rise to a ``nothing-to-something'' tunneling as shown in Fig.(\ref{nts}). This scenario has been previously analyzed in \cite{9903062}.
However, it would be difficult to consider an inhomogeneous instanton in this case. In fact, there are several problems regarding this scenario.
\begin{itemize}
\item[(1)] There is a fixed point (indicated by the blue dots in Fig.(\ref{fixedpoints})) and it is not clear whether this point will give rise to any problem when one calculates the action. Note that this fixed point is ``far away'' from the identification surface where the Euclidean geometry meets the Lorentzian one, so there is no problem at least at the level of pasting the surface to the Lorentzian geometry.  Nevertheless, one might worry that the calculation of the wave function might be ``obstructed'' by the presence of a fixed point\footnote{In the literature, an instanton with a singularity at the south pole is known, and the means to avoid the singularity has been discussed \cite{9803156}. (See, however, \cite{9803084}.)  Though not a singularity, one might be able to treate an orbifold fixed point in a similar way, however we shall not pursue this further in this paper.}. 

\item[(2)] If we do couple a scalar field to gravity, then we have to worry about the boundary condition. If we prescribe the Neumann condition at the blue dashed line ($\phi = 0$), then there may be some non-trivial oscillating solutions (with $\mathbb{Z}_{2}$ symmetry \cite{0410142,1206.7040}). Such a solution has more than one negative mode \cite{coleman, 0602039}. Therefore, this may not be the most preferred path for tunneling.

\item[(3)] Even if there is no scalar field, the probability of the instanton tunneling is much less than that of the original Hawking-Moss instanton. 
To see this, note that the volume of the Euclidean manifold is halved under the antipodal map, and since $\text{Re}(I)$ is negative for de Sitter space, we have the probability $e^{- \text{Re}(I)} <  e^{- 2\text{Re}(I)}$.

\end{itemize}

 \begin{figure}
\includegraphics[width=3.3in]{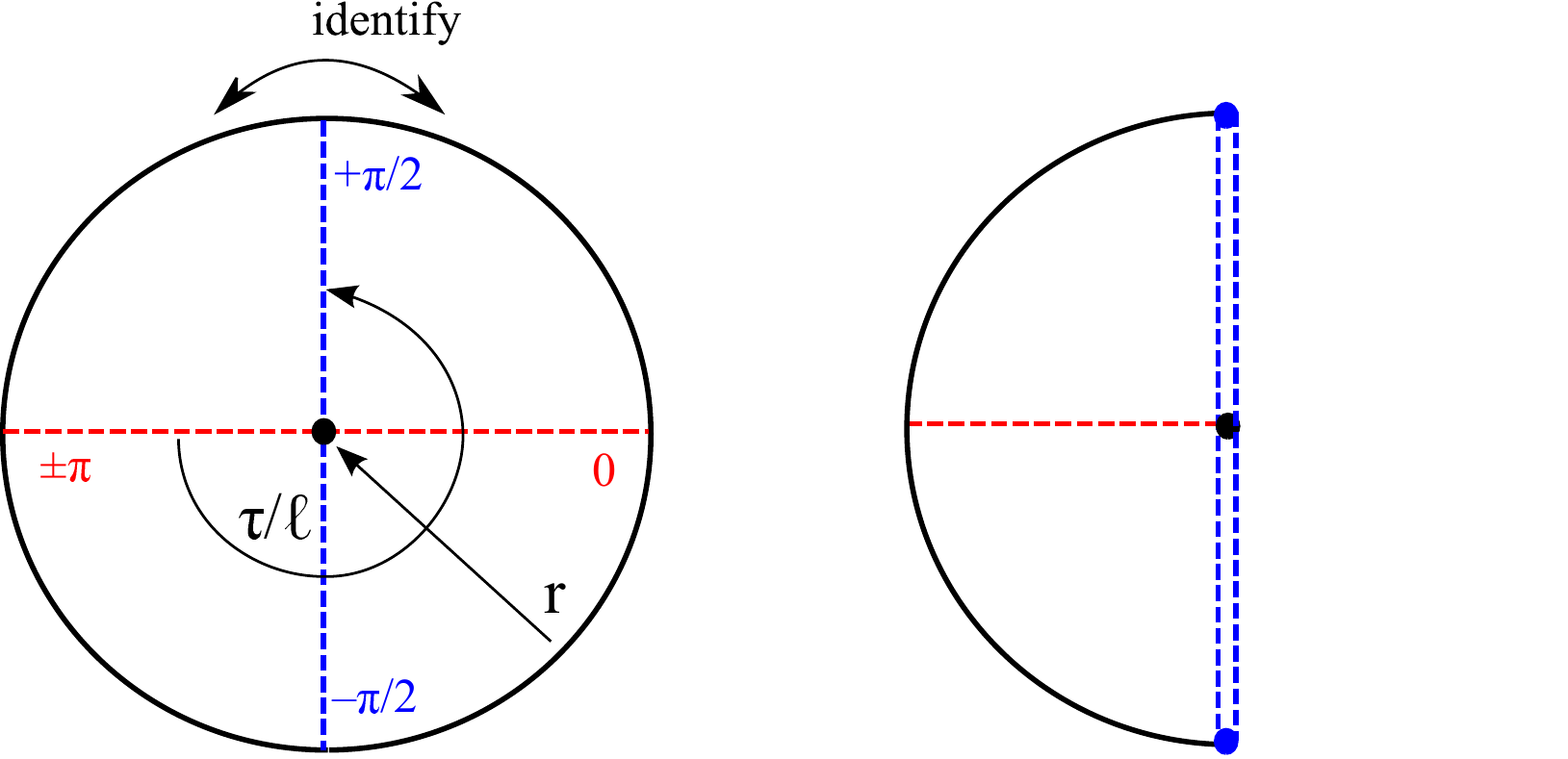} 
\caption{\textbf{Left:} The antipodal map acting on the 4-sphere; here only one of the angular coordinate $\tau/\ell$ is shown. \textbf{Right:} The resulting quotient geometry consists of a hemisphere of $\text{S}^4$, the boundary in blue corresponds to a $\mathbb{R}\text{P}^3$. We have two fixed points indicated by the blue dots. \label{fixedpoints}} 
\end{figure}  

 \begin{figure}
\includegraphics[width=2.0in]{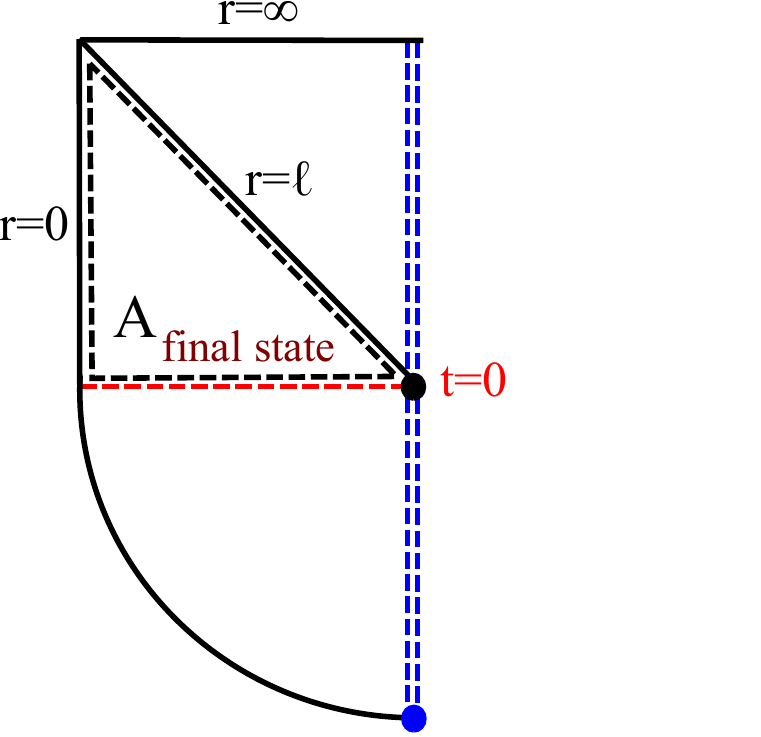} 
\caption{A ``nothing-to-something'' tunneling from $\text{EdS}[\mathbb{R}\text{P}^3]$ to $\text{dS}[\mathbb{R}\text{P}^3]$. The blue dot is a fixed point in the topological identification of the Euclidean geometry.\label{nts}}
\end{figure}  
\end{subsection}

\begin{subsection}{Case III: Something-to-Something with Euclidean Sphere Instanton}

Another possibility is to consider a ``something-to-something'' tunneling, but instead of a Euclidean projective space as the instanton, we just take the good old Euclidean 4-sphere $\text{S}^4$. We illustrate this possibility in Fig.(\ref{noid}), which bears a lot of resemblance with Fig.(\ref{wickrotateRP3}). However, at the matching surface, instead of a red double line that denotes $\mathbb{R}\text{P}^3$ equator, we now have a red single line that denotes the usual, $\text{S}^3$, equator.  Note that except the black dot in the middle which is quite peculiar, every interior point in this diagram corresponds to a $\text{S}^2$. In particular, this means that topologically there is no obstruction at the matching surface \emph{almost everywhere} (in the technical sense of a Lebesgue (volume) measure). The only problem is at the black dot --- there on the Euclidean side it is a $\text{S}^2$, but on the Lorentzian side it is $\mathbb{R}\text{P}^2$, so matching fails at the cosmological horizon. However, it is not clear if this is a serious obstacle, since the problematic region is a set of measure zero. 

 \begin{figure}[h!]
\includegraphics[width=1.8in]{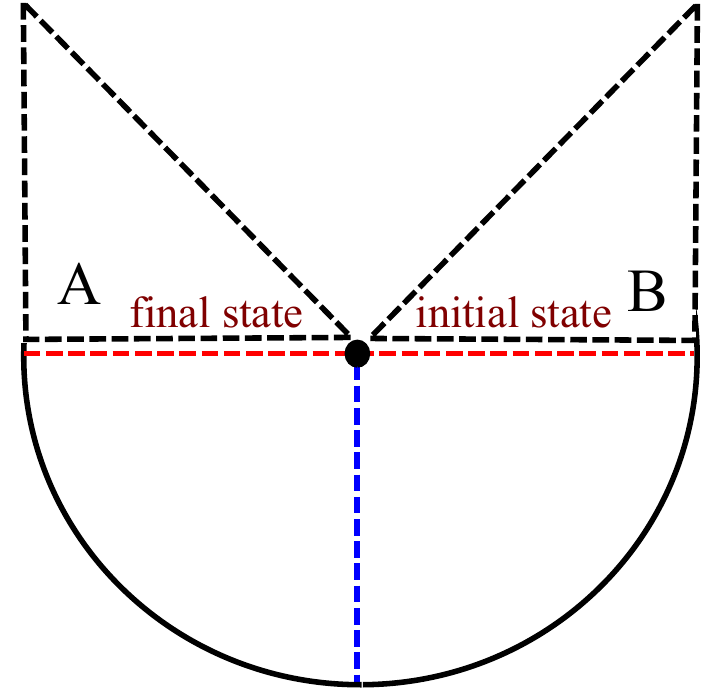} 
\caption{A ``something-to-something'' tunneling from $\text{S}^4$ to $\text{dS}[\mathbb{R}\text{P}^3]$.\label{noid}}
\end{figure}  

Since the instanton is $\text{S}^4$, the probability of tunneling into $\text{dS}[\mathbb{R}\text{P}^3]$ is the same as the tunneling probability from $\text{S}^4$ to $\text{dS}[\text{S}^3]$. Thus Case III, if it works, has tunneling probability that is higher compared to Case II. 

We emphasize that under this something-to-something tunneling scenario, \emph{everything happens in a single causal patch}. It is not about connecting two causally disconnected patches, but an analytic continuation from one hypersurface to another hypersurface, in one single causal patch.

\end{subsection}

\begin{section}{Discussion}
The accelerating expansion of the universe is well-described by the presence of a positive cosmological constant $\Lambda > 0$. Despite many attempts to explain the accelerating expansion using either modified gravity or various forms of dark energy, the cosmological constant remains the simplest explanation that also fits the observational data very well. Our universe is therefore well-modeled by an asymptotically de Sitter space. However, on the theoretical level, the effort to understand asymptotically de Sitter spaces has just barely begun (it is \emph{not} true that since $\Lambda$ is small, one can always assume that the results obtained in asymptotically flat geometries apply straightforwardly, see \cite{1409.3816,1510.05593,1506.06152}), and many puzzles remain to be fully solved \cite{1205.3855}. 

In this work, we are concerned with the global topology of the de Sitter space, which although commonly taken to be $\mathbb{R} \times \text{S}^3$, is \emph{not} what de Sitter originally had in mind. Instead, what de Sitter considered was a spacetime with topology $\mathbb{R} \times \mathbb{R}\text{P}^3$, which we denoted as $\text{dS}[\mathbb{R}\text{P}^3]$. Following \cite{0308022}, we explained why $\text{dS}[\mathbb{R}\text{P}^3]$ is arguably more physical than $\text{dS}[\text{S}^3]$, despite $\mathbb{R}\text{P}^3$ is often thought of as a ``more complicated'' geometry relative to $\text{S}^3$ in modern eyes. Given the nice features of  $\text{dS}[\mathbb{R}\text{P}^3]$ from both the classical perspective and quantum cosmological perspective (as far as entanglement entropy is concerned) \cite{0308022}, we are motivated to explore the quantum cosmological aspect of $\text{dS}[\mathbb{R}\text{P}^3]$ further. 

Our main concern is to study how $\text{dS}[\mathbb{R}\text{P}^3]$ can be created from instanton tunneling. In this work, we found two (potentially) viable scenarios: a ``nothing-to-something'' tunneling in which the Euclidean geometry corresponds to a $\mathbb{Z}_2$-orbifold, and a ``something-to-something'' tunneling in which the Euclidean manifold is $\text{S}^4$. The latter scenario has a higher tunneling probability, however the matching surface between the Euclidean geometry and the Lorentzian geometry fails at a point (where the cosmological horizon meets the $\mathbb{R}\text{P}^2$ equator on the Lorentzian side). This might not be a serious problem since it is a set of measure zero, but a further investigation will be required to check if this is indeed a viable tunneling scenario. The ``nothing-to-something'' tunneling from $\mathbb{Z}_2$-orbifold is not without issue since it has a fixed point at the south pole, which although ``far away'' from the matching surface, might be problematic in the actual computation of the action. However, the fixed point is milder than the singularities in ``singular instantons''. Again, a further study is required to understand whether this tunneling picture works. We ruled out the “something-to-something” tunneling between the $\mathbb{Z}_2$-orbifold to $\text{dS}[\mathbb{R}\text{P}^3]$ since the fixed points are on the ``matching surface'', and furthermore said matching cannot be done at all due to the Euclidean section having different topology than the Lorentzian section.

To conclude, it was shown in \cite{1512.03914} that $\text{dS}[\text{S}^3]$, whose Euclidean geometry admits two different tunneling pictures: ``something-to-something'' and ``nothing-to-something''. Here, we argue that $\text{dS}[\mathbb{R}\text{P}^3]$ might very well also admit these two pictures of tunneling, but there remain some subtleties that deserve further investigations. Other possible implications, e.g., for inflationary scenario \cite{1509.07270}, might also be worth investigating.
 
\end{section}

\begin{acknowledgments}
YCO wishes to thank the Riemann Fellowship during his visit to the Riemann Center for Geometry and Physics, during which part of this work was carried out. He especially thanks Olaf Lechtenfeld for his warm hospitality. YCO is greatful to Nordita for travel support that made this visit possible. 
YCO  also thanks Brett McInnes, Jorma Louko, and Yao-Chieh Hu, for useful discussions. YCO is partially supported by the National Natural Science Foundation of China (NNSFC), while DY is supported by Leung Center for Cosmology and Particle Astrophysics (LeCosPA) of National Taiwan University (103R4000).
\end{acknowledgments}

%%%%%%%%%%%%%%%%%%%%%%%%%%%%%%%%%%%%%%%%%%%%%%%%%%%%%%%%%%%%%%%%%%%%%%%%%%%%%%%%%%%%%%%%%%%%%%%%%%%%%%%%%%%%%%%%%%%%%%%%%%%%%%%%%%%%%%%%%%%%%%%%%%%%%%%%%%

%\end{multicols}


\begin{thebibliography}{99}

\bibitem{1002.3966}
Eugenio Bianchi, Carlo Rovelli, ``Why All These Prejudices Against a Constant?'', \href{https://arxiv.org/abs/1002.3966}{[arXiv:1002.3966 [astro-ph.CO]]}.

\bibitem{deSitter} Willem de Sitter, ``A. Einstein’s Theory of Gravitation and its Astronomical Consequences'',
Third Paper, {\changeurlcolor{vividviolet}\href{http://mnras.oxfordjournals.org/content/78/1/3.full.pdf+html}{MNRAS \textbf{78} (1917) 3}}. 

\bibitem{0308022}
Brett McInnes, ``De Sitter and Schwarzschild-de Sitter according to Schwarzschild and de Sitter'', {\changeurlcolor{vividviolet}\href{https://iopscience.iop.org/article/10.1088/1126-6708/2003/09/009/meta;jsessionid=5D49C5B3EDE57DCB280D2DA6C1042F78.c1}{JHEP \textbf{0309} (2003) 009}}, \href{http://arxiv.org/abs/hep-th/0308022}{[arXiv:hep-th/0308022]}.

\bibitem{Schwarzschild}
Karl Schwarzschild, ``\"Uber das zul\"assige Kr\"ummungsmass des Raumes'', Vierteljahrschrift
d. Astronom. Gesellschaft \textbf{35} (1900) 337.

\bibitem{Sch}
Karl Schwarzschild, ``On the Permissible Curvature of Space'', {\changeurlcolor{vividviolet}\href{https://iopscience.iop.org/article/10.1088/0264-9381/15/9/003/pdf}{Class. Quantum Grav. \textbf{15} (1998) 2539}}.

\bibitem{1303.5086}
Planck Collaboration, ``Planck 2013 Results. XXVI. Background Geometry and Topology of the Universe'', {\changeurlcolor{vividviolet}\href{http://www.aanda.org/articles/aa/abs/2014/11/aa21546-13/aa21546-13.html}{Astron. Astrophys. \textbf{571} (2014) A26}}, \href{https://arxiv.org/abs/1303.5086}{[arXiv:1303.5086 [astro-ph.CO]]}.

\bibitem{1601.03884}
Jean-Pierre Luminet, ``The Status of Cosmic Topology after Planck Data'', {\changeurlcolor{vividviolet}\href{http://www.mdpi.com/2218-1997/2/1/1}{Universe \textbf{2} (2016) 1}}, \href{http://arxiv.org/abs/1601.03884}{[arXiv:1601.03884 [astro-ph.CO]]}. 

\bibitem{1604.02179}
German I. Gomero, Bruno Mota, Marcelo J. Reboucas, ``Limits of the Circles-in-the-Sky Searches in the Determination of Cosmic Topology of Nearly Flat Universe'', {\changeurlcolor{vividviolet}\href{https://journals.aps.org/prd/abstract/10.1103/PhysRevD.94.043501}{Phys. Rev. D \textbf{94} (2016) 043501}}, \href{http://arxiv.org/abs/1604.02179}{[arXiv:1604.02179 [astro-ph.CO]]}.

\bibitem{9812056}
Jorma Louko, Kristin Schleich, ``The Exponential Law: Monopole Detectors, Bogoliubov Transformations, and the Thermal Nature of the Euclidean Vacuum in $\mathbb{RP}^3$ de Sitter Spacetime'', {\changeurlcolor{vividviolet}\href{https://iopscience.iop.org/article/10.1088/0264-9381/16/6/328/meta}{Class. Quant. Grav. \textbf{16} (1999) 2005}}, \href{http://arxiv.org/abs/gr-qc/9812056}{[arXiv:gr-qc/9812056]}.

\bibitem{0510049}
Paul Langlois, ``Causal Particle Detectors and Topology'', {\changeurlcolor{vividviolet}\href{http://www.sciencedirect.com/science/article/pii/S0003491606000480}{Annals Phys. \textbf{321} (2006) 2027-2070}}, \href{https://arxiv.org/abs/gr-qc/0510049}{[arXiv:gr-qc/0510049]}.

\bibitem{GH}
Garry W. Gibbons, Stephen W. Hawking, ``Cosmological Event Horizons, Thermodynamics and
Particle Creation'', {\changeurlcolor{vividviolet}\href{http://journals.aps.org/prd/abstract/10.1103/PhysRevD.15.2738}{Phys. Rev. D \textbf{15} (1977) 2738}}.

\bibitem{1104.3712}
Sergey N. Solodukhin, ``Entanglement Entropy of Black Holes'', 	{\changeurlcolor{vividviolet}\href{http://relativity.livingreviews.org/Articles/lrr-2011-8/}{Living Rev. Relativity \textbf{14} (2011) 8}}, \href{http://arxiv.org/abs/1104.3712}{[arXiv:1104.3712 [hep-th]]}.

\bibitem{9409015}
Jacob D. Bekenstein, ``Do We Understand Black Hole Entropy?'', \href{https://arxiv.org/abs/gr-qc/9409015}{[arXiv:gr-qc/9409015]}.

\bibitem{9906031v1}
Jorma Louko, ``Single-Exterior Black Holes'', {\changeurlcolor{vividviolet}\href{https://link.springer.com/chapter/10.1007/3-540-46634-7_8}{Lect. Notes Phys. \textbf{541} (2000) 188}}, \href{https://arxiv.org/abs/gr-qc/9906031v1}{[arXiv:gr-qc/9906031]}.

\bibitem{1001.0124v1}
Jorma Louko, ``Geon Black Holes and Quantum Field Theory'', {\changeurlcolor{vividviolet}\href{https://iopscience.iop.org/article/10.1088/1742-6596/222/1/012038/meta;jsessionid=262807BEF651ED857CD6F35011F908BB.c4.iopscience.cld.iop.org}{J. Phys. Conf. Ser. \textbf{222} (2010) 012038}}, \href{https://arxiv.org/abs/1001.0124v1}{[arXiv:1001.0124 [gr-qc]]}.

\bibitem{1601.03447}
Gerard 't Hooft, ``Black Hole Unitarity and Antipodal Entanglement'', {\changeurlcolor{vividviolet}\href{https://link.springer.com/article/10.1007\%2Fs10701-016-0014-y}{Found. Phys. \textbf{46} (2016) 1185}}, \href{https://arxiv.org/abs/1601.03447}{[arXiv:1601.03447 [gr-qc]]}.

\bibitem{DeWitt}
Bryce S. DeWitt, ``Quantum Theory of Gravity. I. The Canonical Theory'', {\changeurlcolor{vividviolet}\href{http://journals.aps.org/pr/abstract/10.1103/PhysRev.160.1113}{Phys. Rev. 160 (1967) 1113}}.

\bibitem{FGG}
Edward Farhi, Alan H. Guth, Jemal Guven, ``Is it Possible to Create a Universe in the Laboratory by Quantum Tunneling?'', {\changeurlcolor{vividviolet}\href{http://www.sciencedirect.com/science/article/pii/055032139090357J}{Nucl. Phys. B \textbf{339} (1990) 417}}.

\bibitem{FMP1}
Willy Fischler, Daniel Morgan, Joseph Polchinski, ``Quantum Nucleation of False-Vacuum Bubbles'', {\changeurlcolor{vividviolet}\href{http://journals.aps.org/prd/abstract/10.1103/PhysRevD.41.2638}{Phys. Rev. D \textbf{41} (1990) 2638(R)}}.

\bibitem{FMP2}
Willy Fischler, Daniel Morgan, Joseph Polchinski, ``Quantization of False-Vacuum Bubbles: A Hamiltonian Treatment of Gravitational Tunneling'', {\changeurlcolor{vividviolet}\href{http://journals.aps.org/prd/abstract/10.1103/PhysRevD.42.4042}{Phys. Rev. D \textbf{42} (1990) 4042}}.

\bibitem{HH}
James Hartle, Stephen Hawking, ``Wave Function of the Universe''. {\changeurlcolor{vividviolet}\href{http://journals.aps.org/prd/abstract/10.1103/PhysRevD.28.2960}{Phys. Rev. D \textbf{28} (1983) 2960}}.

\bibitem{1512.03914}
Pisin Chen, Yao-Chieh Hu, Dong-han Yeom, ``Two Interpretations on Thin-Shell Instantons'', {\changeurlcolor{vividviolet}\href{https://journals.aps.org/prd/abstract/10.1103/PhysRevD.94.024044}{Phys. Rev. D \textbf{94} (2016) 024044}}, \href{http://arxiv.org/abs/1512.03914}{[arXiv:1512.03914 [hep-th]]}.

\bibitem{1110.0611v1}
Gary W. Gibbons, ``Topology Change in Classical and Quantum Gravity'', \href{https://arxiv.org/abs/1110.0611v1}{[arXiv:1110.0611 [gr-qc]]}.

\bibitem{CDL}
Sidney Coleman, Frank De Luccia, ``Gravitational Effects on and of Vacuum Decay'', {\changeurlcolor{vividviolet}\href{http://journals.aps.org/prd/abstract/10.1103/PhysRevD.21.3305}{Phys. Rev. D \textbf{21} (1980) 3305}}.

\bibitem{HM}
Stephen W. Hawking, Ian L. Moss, ``Supercooled Phase Transitions in the Very Early Universe'',  {\changeurlcolor{vividviolet}\href{http://www.sciencedirect.com/science/article/pii/0370269382909467}{Phys. Lett. B \textbf{110} (1982) 35}}.

\bibitem{9802030}
Stephen W. Hawking, Neil Turok, ``Open Inflation Without False Vacua'', {\changeurlcolor{vividviolet}\href{http://www.sciencedirect.com/science/article/pii/S0370269398002342}{Phys. Lett. B \textbf{425} (1998) 25}}, \href{https://arxiv.org/abs/hep-th/9802030}{[arXiv:hep-th/9802030]}.

\bibitem{0706.1573}
Adam R. Brown, Erick J. Weinberg, ``Thermal Derivation of the Coleman-De Luccia Tunneling Prescription'', {\changeurlcolor{vividviolet}\href{http://journals.aps.org/prd/abstract/10.1103/PhysRevD.76.064003}{Phys. Rev. D \textbf{76} (2007) 064003}}, \href{https://arxiv.org/abs/0706.1573}{[arXiv:0706.1573 [hep-th]]}.

\bibitem{visser}
Matt Visser, \emph{Lorentzian Wormholes from Einstein to Hawking}, AIP Press, 1995.

\bibitem{9607079v2}
Andrew Chamblin, Gary W. Gibbons, ``Nucleating Black Holes via Non-Orientable Instantons'', {\changeurlcolor{vividviolet}\href{http://journals.aps.org/prd/abstract/10.1103/PhysRevD.55.2177}{Phys. Rev. D \textbf{55} (1997) 2177}}, \href{https://arxiv.org/abs/gr-qc/9607079v2}{[arXiv:gr-qc/9607079]}.

\bibitem{1603.04426}
Alexander Maloney, Simon F. Ross, ``Holography on Non-Orientable Surfaces'', {\changeurlcolor{vividviolet}\href{https://iopscience.iop.org/0264-9381/33/18/185006/}{Class. Quant. Grav. \textbf{33} (2016) 185006}},\href{https://arxiv.org/abs/1603.04426}{[arXiv:1603.04426 [hep-th]]}.


\bibitem{Ratcliffe}
John G. Ratcliffe, Steven T. Tschant, ``Gravitational Instantons of Constant Curvature'', {\changeurlcolor{vividviolet}\href{https://iopscience.iop.org/article/10.1088/0264-9381/15/9/009/meta}{Class. Quant. Gravit \textbf{15} (1998) 2613}}.

\bibitem{9805101}
Alan Daughton, Jorma Louko, Rafael D. Sorkin, ``Instantons and Unitarity in Quantum Cosmology with Fixed Four-Volume'', {\changeurlcolor{vividviolet}\href{https://journals.aps.org/prd/abstract/10.1103/PhysRevD.58.084008}{Phys. Rev. D \textbf{58} (1998) 084008}}, \href{https://arxiv.org/abs/gr-qc/9805101}{[arXiv:gr-qc/9805101]}.


\bibitem{9505035}
John L. Friedman, Atsushi Higuchi, ``Quantum Field Theory in Lorentzian Universes-From-Nothing'', {\changeurlcolor{vividviolet}\href{https://journals.aps.org/prd/abstract/10.1103/PhysRevD.52.5687}{Phys. Rev. D \textbf{52} (1995) 5687}}, \href{https://arxiv.org/abs/gr-qc/9505035}{[arXiv:gr-qc/9505035]}.

\bibitem{9803156}
Neil Turok, Stephen W. Hawking, ``Open Inflation, the Four Form and the Cosmological Constant'', {\changeurlcolor{vividviolet}\href{http://www.sciencedirect.com/science/article/pii/S0370269398006510}{Phys. Lett. B \textbf{432} (1998) 271}}, \href{https://arxiv.org/abs/hep-th/9803156}{[arXiv:hep-th/9803156]}.

\bibitem{9803084}
Alexander Vilenkin, ``Singular Instantons and Creation of Open Universes'', {\changeurlcolor{vividviolet}\href{https://journals.aps.org/prd/abstract/10.1103/PhysRevD.57.R7069}{Phys. Rev. D \textbf{57} (1998) 7069}}, \href{https://arxiv.org/abs/hep-th/9803084}{[arXiv:hep-th/9803084]}.

\bibitem{9903062}
Kristin Schleich, Donald M. Witt, ``The Generalized Hartle-Hawking Initial State: Quantum Field Theory on Einstein Conifolds'', {\changeurlcolor{vividviolet}\href{https://journals.aps.org/prd/abstract/10.1103/PhysRevD.60.064013}{Phys. Rev. D \textbf{60} (1999) 064013}}, \href{https://arxiv.org/abs/gr-qc/9903062}{[arXiv:gr-qc/9903062]}.

\bibitem{0410142}
James C. Hackworth, Erick J. Weinberg, ``Oscillating Bounce Solutions and Vacuum Tunneling in de Sitter Spacetime'', {\changeurlcolor{vividviolet}\href{https://journals.aps.org/prd/abstract/10.1103/PhysRevD.71.044014}{Phys. Rev. D \textbf{71} (2005) 044014}}, \href{https://arxiv.org/abs/hep-th/0410142}{[arXiv:hep-th/0410142]}.

\bibitem{1206.7040}
Bum-Hoon Lee, Wonwoo Lee, Dong-han Yeom, ''Oscillating Instantons As Homogeneous Tunneling Channels'', {\changeurlcolor{vividviolet}\href{http://www.worldscientific.com/doi/abs/10.1142/S0217751X13500826}{Int. J. Mod. Phys. A \textbf{28} (2013) 1350082}}, \href{https://arxiv.org/abs/1206.7040}{[arXiv:1206.7040 [hep-th]]}.

\bibitem{coleman}
Sidney Coleman, ``Quantum Tunneling and Negative Eigenvalues'', {\changeurlcolor{vividviolet}\href{http://www.sciencedirect.com/science/article/pii/0550321388903082}{Nucl. Phys. B \textbf{298} (1988) 178}}.

\bibitem{0602039}
George Lavrelashvili, ``The Number of Negative Modes of the Oscillating Bounces'',  {\changeurlcolor{vividviolet}\href{https://journals.aps.org/prd/abstract/10.1103/PhysRevD.73.083513}{Phys. Rev. D \textbf{73} (2006) 083513}}, \href{https://arxiv.org/abs/gr-qc/0602039}{[arXiv:gr-qc/0602039]}.



\bibitem{1409.3816}
Abhay Ashtekar, B{\'e}atrice Bonga, Aruna Kesavan, ``Asymptotics with a Positive Cosmological Constant: I. Basic Framework'', {\changeurlcolor{vividviolet}\href{http://iopscience.iop.org/article/10.1088/0264-9381/32/2/025004/meta;jsessionid=94F9A0CBB1AAF385AF975EF344FF8273.c4.iopscience.cld.iop.org}{Class. Quant. Grav. \textbf{32} (2015) 025004}}, \href{https://arxiv.org/abs/1409.3816}{[arXiv:1409.3816 [gr-qc]]}.

\bibitem{1506.06152}
Abhay Ashtekar, B{\'e}atrice Bonga, Aruna Kesavan, ``Asymptotics with a Positive Cosmological Constant: II. Linear fields on de Sitter Space-time'', {\changeurlcolor{vividviolet}\href{https://journals.aps.org/prd/abstract/10.1103/PhysRevD.92.044011}{Phys. Rev. D \textbf{92} (2015) 044011}}, \href{https://arxiv.org/abs/1506.06152}{[arXiv:1506.06152 [gr-qc]]}.

\bibitem{1510.05593}
Abhay Ashtekar, B{\'e}atrice Bonga, Aruna Kesavan, ``Asymptotics with a Positive Cosmological Constant: III. The Quadrupole Formula'', {\changeurlcolor{vividviolet}\href{https://journals.aps.org/prd/abstract/10.1103/PhysRevD.92.104032}{Phys. Rev. D \textbf{92} (2015) 10432}}, \href{https://arxiv.org/abs/1510.05593}{[arXiv:1510.05593 [gr-qc]]}.

\bibitem{1205.3855}
Dionysios Anninos, ``De Sitter Musings'', {\changeurlcolor{vividviolet}\href{http://www.worldscientific.com/doi/abs/10.1142/S0217751X1230013X}{Int. J. Mod. Phys. A \textbf{27} (2012) 1230013}}, \href{https://arxiv.org/abs/1205.3855}{[arXiv:1205.3855 [hep-th]]}. 

\bibitem{1509.07270}
Andrei O. Barvinsky, Aleksander Yu. Kamenshchik, Dmitry V. Nesterov, ``New Type of Hill-Top Inflation'', {\changeurlcolor{vividviolet}\href{https://iopscience.iop.org/1475-7516/2016/01/036/}{JCAP \textbf{1601} (2016) 036}}, \href{https://arxiv.org/abs/1509.07270}{[arXiv:1509.07270 [hep-th]]}.

%

\end{thebibliography}
\end{document}